\documentclass[aps,prb,twocolumn,superscriptaddress,longbibliography, nofootinbib]{revtex4-2}
\usepackage[T1]{fontenc}
\usepackage[utf8]{inputenc}
\usepackage{color}
\usepackage{xcolor}
\usepackage{graphicx}
\usepackage[english]{babel}
\usepackage{amsmath,amssymb}
\usepackage{dsfont}
\usepackage{physics}
\usepackage{quantikz}
\usepackage[unicode=true,pdfusetitle,
bookmarks=true,bookmarksnumbered=false,bookmarksopen=false,
 breaklinks=true,pdfborder={0 0 0},pdfborderstyle={},backref=false,colorlinks=true]
 {hyperref}
\usepackage[printonlyused,withpage,nolist]{acronym}

\newcommand{\eviden}{Eviden Quantum Lab, 78340 Les Clayes-sous-Bois, France}
\newcommand{\CDF}{Coll\`ege de France, Universit\'e PSL, 11 place Marcelin Berthelot, 75005 Paris, France}
\newcommand{\CPHT}{CPHT, CNRS, Ecole Polytechnique, Institut Polytechnique de Paris, 91128 Palaiseau, France}

\newcommand{\nodag}{{\vphantom{\dagger}}}
\newcommand{\systS}{\mathcal{S}}
\newcommand{\systB}{\mathcal{B}}
\newcommand{\systE}{\mathcal{E}}
\newcommand{\systA}{\mathcal{A}}
\newcommand{\altlindblad}{\underline{\mathcal{L}}}
\newcommand{\identity}{\mathds{1}}

\begin{document}
\title{
Turning qubit noise into an advantage: automatic state preparation and long-time dynamics for impurity models on quantum computers
}

\author{Corentin Bertrand}
\email[]{corentin.bertrand@eviden.com}
\affiliation{\eviden}
\author{Pauline Besserve}
\affiliation{\eviden}
\affiliation{\CDF}
\affiliation{\CPHT}
\author{Michel Ferrero}
\affiliation{\CDF}
\affiliation{\CPHT}
\author{Thomas Ayral}
\affiliation{\eviden}

		
\begin{abstract}
Noise is often regarded as a limitation of quantum computers. 
In this work, we show that in the dynamical mean field theory (DMFT) approach to strongly-correlated systems, it can actually be harnessed to our advantage. 
Indeed, DMFT maps a lattice model onto an impurity model, namely a finite system coupled to a dissipative bath.
While standard approaches require a large number of high-quality qubits in a unitary context, 
we propose a circuit that harvests amplitude damping to reproduce the dynamics of this model with a blend of noisy and noiseless qubits.
We find compelling advantages with this approach: a substantial reduction in the number of qubits, the ability to reach longer time dynamics, and no need for ground state search and preparation.
This method would naturally fit in a partial quantum error correction framework.
\end{abstract}
\maketitle


\begin{acronym}
    \acro{GF}{Green's function}
    \acro{HF}{hybridization function}
    \acro{DMFT}{dynamical mean field theory}
    \acro{PSD}{positive semi definite}
    \acro{RLM}{resonant level model}
    \acro{1-RDM}{1-particle reduced density matrix}
    \acro{NISQ}{near-term intermediate scale quantum}
    \acro{QRT}{quantum regression theorem}
    \acro{JW}{Jordan-Wigner}
    \acro{PM}{pseudomode}
    \acro{CZ}{controlled-$Z$}
\end{acronym}

\acresetall

One of the most promising applications of quantum computing is the study of quantum many-body systems~\cite{Ayral2023}.
More often than not, systems found in Nature are coupled to an environment, so that the relevant dynamics is 
not unitary.
This is the case for instance of
electrons in strongly correlated materials~\cite{tokura_emergent_2017}, which are immersed in the thermodynamical bath formed by the rest of the material.
Yet, quantum computing is usually framed in terms of unitary operations,
and an overarching goal of today's hardware and software efforts is to enforce this unitarity through
noise reduction, mitigation~\cite{Cai2022} or correction~\cite{Terhal2015}.
Ironically, even if these efforts were to bear fruit, simulating dissipative dynamics would still require large amounts of qubits to represent the many degrees of freedom of a thermodynamical bath.
Why not, then, use quantum noise as a direct source of dissipation?

While dissipative quantum computing has already been explored for quantum state preparation, entanglement stabilization~\cite{pocklington_stabilizing_2022, harrington_engineered_2022}, optimization tasks~\cite{chen_local_2023} and representation of bosonic baths~\cite{leppakangas_quantum_2023, rost_simulation_2020, sun_efficient_2021}, here we explore its potential for simulating electrons in strongly correlated materials.
The dissipative nature of electrons in materials is made explicit by \ac{DMFT}~\cite{Georges1996}, one of the most powerful methods to describe exotic phases of  matter. It maps a lattice model onto a self-consistent impurity model, which describes a finite interacting system (the impurity, typically one or a few atomic sites) coupled to a noninteracting, non-Markovian bath (effectively representing the rest of the lattice).
The bottleneck of \ac{DMFT} lies in computing the impurity model's dynamics.
Only a handful of classical algorithms are known to provide controlled solutions: \textit{e.g. }
different flavors of quantum Monte-Carlo algorithms, limited by the sign problem~\cite{gull_continuous-time_2011, simons_collaboration_2015}, and tensor network algorithms, which struggle with complex entanglement structures~\cite{wolf_solving_2014, bauernfeind_fork_2017}.
This has prompted efforts to use unitary quantum processors~\cite{Kreula2016,Kreula2016a,Bauer2016,Rungger2019,Yao2020,Jaderberg2020, Besserve2021} to tackle these problems, with an elephant in the room: the large number of qubits, and the ensuing deep circuits, required to deal with thermodynamical baths.

\begin{figure*}
    \centering
    \includegraphics[width=1.\textwidth]{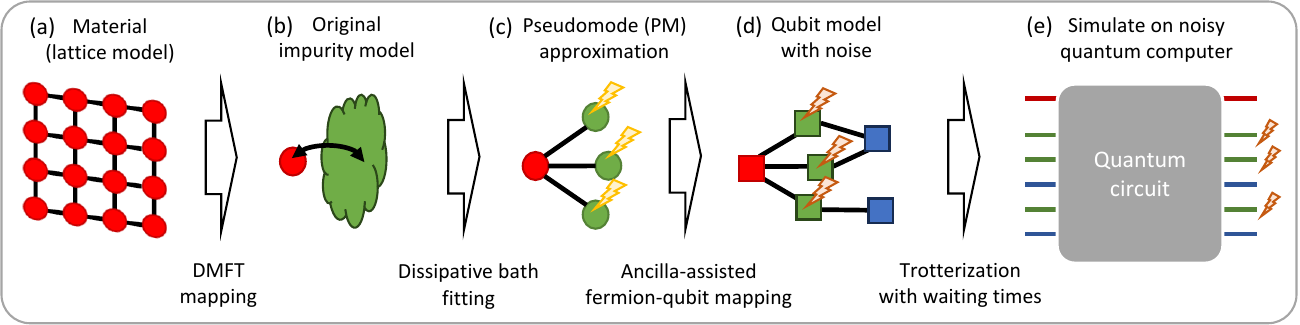}
    \caption{\textit{Summary:} A lattice problem (a) is mapped onto an atomic problem (red dot) coupled to a bath (green cloud) (b), which is itself represented by dissipative fermionic modes (green dots) through a fit of the hybridization function (c). Through a fermion-qubit mapping using ancillas, we obtain a noiseless qubit (red square) coupled to noisy and noiseless qubits (green and blue squares) (d). The time evolution needed to compute e.g GFs is realized using a quantum circuit (e).}
    \label{fig:hybrid_architecture}
\end{figure*}

In this work, we propose a method to use intrinsic qubit noise, and specifically amplitude damping, to reproduce the effect of a structured fermionic bath.
It extends the approach of Ref~\cite{leppakangas_quantum_2023} to fermions.
We first build a compact representation of the bath with \acp{PM}~\cite{garraway_nonperturbative_1997}, i.e. an open system exchanging fermions with a Markovian environment (Fig.~\ref{fig:hybrid_architecture} (b) to (c)).
We then describe a fermion-to-qubit transformation mapping fermionic dissipative processes onto qubit amplitude damping (step (c) to (d)).
Assuming the availability of a few noiseless qubits and many noisy qubits, we provide a quantum circuit that harvests qubit noise to reproduce the original model dynamics (step (d) to (e)).
We show that taking advantage of noise not only allows to reach longer time dynamics with fewer qubits, but also provides automatic state preparation, as noise naturally drives the system to its steady state.

Our algorithm could readily be implemented on hardware that is mainly sensitive to amplitude damping, such as $T_1$-limited superconducting qubits~\cite{kakuyanagi_dephasing_2007, kubica_erasure_2023}.
On more standard superconducting qubit architectures, pure dephasing is important and would have to be mitigated, using e.g. zero-noise extrapolation~\cite{temme_error_2017}.
As we use a blend of clean and noisy qubits, our scheme naturally fits in a partial quantum error correction framework~\cite{bultrini_battle_2023, koukoulekidis_framework_2023}.

\newpage

\section{Model}
We focus on impurity models, 
defined by a Keldysh action~\cite{kamenev2023}:
\begin{gather}
\label{eq:action}
    \mathcal{S} = \int_{\mathcal{C}} \dd{t} H_{\rm imp}[\vb{d}^\dag(t), \vb{d}(t)] + \iint_{\mathcal{C}} \dd{t}\dd{t'} \vb{d}^\dag(t) \cdot \Delta(t, t') \cdot \vb{d}(t')
    \\
    H_{\rm imp}[\vb{d}^\dag, \vb{d}] = \sum_{ij} h_{ij} d_i^* d_j + \frac{1}{2} \sum_{ijkl} v_{ijkl} d_i^* d_j^* d_k d_l
\end{gather}
where $\vb{d}^\dag = (d^*_1, d^*_2, \ldots)$ (resp. 
$\vb{d} = (d_1, d_2, \ldots)$) are two independent line (column) vectors of Grassmann fields representing the impurity fermionic modes (indices may include spin), $H_{\rm imp}$ is the impurity Hamiltonian with hopping energies $h_{ij}$ and interaction energies $v_{ijkl}$.
$\Delta(t, t')$ is the bath \ac{HF} matrix, evaluated on the Keldysh contour $\mathcal{C}$~\cite{kamenev2023}.
Eq.~\eqref{eq:action} highlights the crucial fact that the bath, being noninteracting, influences the impurity's dynamics only through its \ac{HF} $\Delta$.
Solving Eq.~\eqref{eq:action} amounts to computing the impurity's one-body \acp{GF}, for example
the greater \ac{GF} between impurity modes $i$ and $j$, defined by the path integral
\begin{equation}
    G_{ij}^>(t) = -i \int \mathcal{D}[\vb{d}^\dag, \vb{d}] \; d_i^\nodag(t + t_{\rm prep}) d_j^*(t_{\rm prep}) e^{i\mathcal{S}[\vb{d}^\dag, \vb{d}]},
\end{equation}
which is independent of $t_{\rm prep}$ in the steady-state limit.
In all numerical examples below, we illustrate our method for $H_{\rm imp} = \varepsilon d^\dag d$ (called the \ac{RLM}~\cite{emery_mapping_1992}), that describes a single, spinless, free ($v=0$) fermionic level in an infinite, half-filled thermal bath (of temperature $1/\beta = 1$, chosen as the energy unit), with a hybridization spectral function
$\Delta(\omega) = \gamma^2 / (2 D)$ if $|\omega| \le D$ and $=0$ elsewhere, where $D=10/\beta$ is the half bandwidth and $\gamma=0.6/\beta$ the impurity-bath coupling energy. We pick $\varepsilon = 0.5 /\beta$. The lesser and greater \acp{HF} thus read $  \Delta^{\gtrless}(\omega) = \mp\frac{2\pi i}{1 + e^{\mp\beta\omega}} \Delta(\omega)$.
This simple case will allow us to provide exact results, but the methodology below is valid for the interacting ($v\neq 0$) and time-dependent ($H_\mathrm{imp}(t)$) cases, and for any \acp{HF}.

Even though we focus on the \ac{RLM}, models with several fermion flavors can be treated similarly.
For example, in the case of the Anderson impurity model, where fermions have spin, one can apply the bath pseudomode representation on each spin species separately.

\section{Bath representation with pseudomodes}
The first step of our method is to build a compact representation of the bath, i.e. with as few fermionic modes as possible, in order to limit the number of qubits and gates.

In standard \ac{DMFT}~\cite{wolf_solving_2014, gramsch_hamiltonian-based_2013}, one uses a closed system to reproduce a given \ac{HF}.
This is done by discretizing frequencies and associating each frequency $\epsilon_{p}$ to a fermionic "bath" mode.
Indeed, the effect of linearly coupling the impurity to a noninteracting bath is completely characterized by the lesser and greater components of the \ac{HF}, $\Delta^{\gtrless}(\omega)$~\cite{tamascelli_nonperturbative_2018, kamenev2023}, which can be regarded as the bath's emission and absorption spectra
(App.~\ref{app:theory}).
In this approach, the initial state must be prepared in the ground state, or in a Gibbs state for nonzero temperatures.

\begin{figure}[t]
    \centering
    \includegraphics[width=\linewidth]{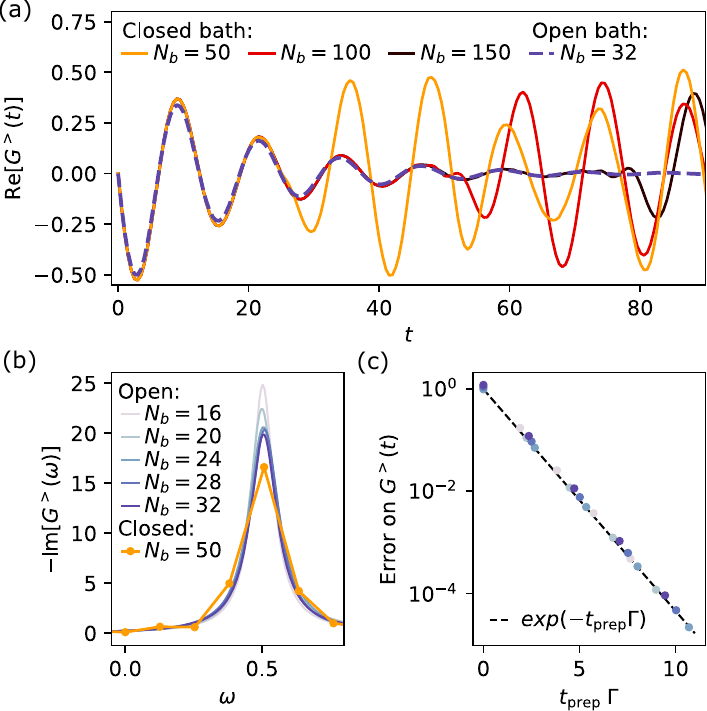}
    \caption{
        \label{fig:noisy_vs_noiseless}
        \textit{Open vs. closed bath representation.}
        (a) \ac{GF} in time domain obtained with large closed baths ($N_{\rm b} \ge 50$ plain lines, warm colors), and with a small open bath ($N_{\rm b} = 32$, dashed blue line).
        (b) Emission spectral density obtained with a large closed bath ($N_{\rm b} = 50$, yellow) and small open baths ($N_{\rm b} \le 32$, blue lines).
        (c) Error on \ac{GF} obtained with open baths of different sizes (dots, shades of blue) with state preparation time $t_\mathrm{prep}$.
    }
\end{figure}

However, this representation of the \acp{HF} with a finite and closed system can only be faithful at short times.
Instead, the \acf{PM} method---a widespread technique to describe non-Markovian environments by master equations both in a bosonic ~\cite{garraway_nonperturbative_1997, xu_theory_2017, tamascelli_nonperturbative_2018, lambert_modelling_2019, pleasance_generalized_2020, pleasance_pseudomode_2021, chen_bexcitonics_2024, menczel_non-hermitian_2024, lacroix_description_2015, lacroix_non-markovian_2020, park_quasi-lindblad_2024} and in a fermionic~\cite{arrigoni_nonequilibrium_2013, dorda_auxiliary_2014, dorda_auxiliary_2015, schwarz_lindblad-driven_2016, dorda_optimized_2017, fugger_nonequilibrium_2018, schwarz_nonequilibrium_2018, chen_markovian_2019, chen_auxiliary_2019, sorantin_auxiliary_2019, fugger_nonequilibrium_2020, lotem_renormalized_2020, wojtowicz_open_2020, elenewski_performance_2021, cirio_pseudofermion_2023} context---couples each bath mode to a Markovian environment that generates
dissipative dynamics.
One major advantage of this approach is that it requires fewer bath modes than the closed system approach.
It is described by a Lindblad equation
\begin{equation}
\label{eq:Lindblad}
    \frac{d\rho}{dt} = -i\qty[H,\rho] + \sum_{p=0}^{N_{\rm b}-1}\qty(2 L_{p}\rho L_{p}^{\dagger} - \qty{ L_{p}^{\dagger}L_{p},\rho} ), 
\end{equation}
with a Hamiltonian $H = H_{\rm imp} + \sum_{p=0}^{N_{\rm b}-1} V_{p} \qty(c^\dag_p d^\nodag + d^\dag c^\nodag_p) + \epsilon_p c^\dag_p c^\nodag_p$. Here, $c_p^\dag$ (resp. $c_p^\nodag$) creates (annihilates) an electron on the $p$th bath mode.
Bath sites are split between \emph{emitters} (even index) and \emph{absorbers} (odd index) and associated to  jump operators
$L_{2k} =\sqrt{\Lambda} c_{2k}^{\dagger}$ and
$L_{2k+1} =\sqrt{\Lambda} c_{2k+1}$.
In this work, without loss of generality, we impose an equal jump rate $\Lambda$ across all bath modes.

The real parameters $V_{p}$, $\epsilon_p$ and $\Lambda$ are optimized so that the \ac{HF} in the \ac{PM} model, $\Delta_{\rm PM}$,
whose steady-state retarded component reads
(App.~\ref{sec:diss_hybridization})
\begin{equation}
    \Delta_{\rm PM}^R(\omega) = \sum_{p=0}^{N_{\rm b}-1} \frac{V_p^2} {\omega - \epsilon_p + i \Lambda},\label{eq:diss_hyb}
\end{equation}
fits the original \ac{HF}, $\Delta$, of Eq.~\eqref{eq:action} as closely as possible
(See details in App.~\ref{app:fitting}).
Intuitively, Eq.~\eqref{eq:diss_hyb} reveals that by using dissipative bath modes, we fit the \ac{HF} with (a few) Lorentzian functions of width $\Lambda$ instead of (a lot of) Dirac peaks.

This leads to the first advantage of the \ac{PM} approach: it unlocks long-time dynamics.
With a closed bath of $N_{\rm b}$ sites, the impurity \acp{GF} witness a revival effect after a time proportional to $N_{\rm b}$~\cite{de_vega_how_2015}.
Long-time dynamics, which is important with e.g. weakly coupled baths such as in Kondo physics~\cite{hewson_kondo_1993}, therefore requires many bath sites.
For instance, in our \ac{RLM}, we need no fewer than $N_{\rm b} \approx 150$ bath sites to capture the whole relaxation dynamics (Fig.~\ref{fig:noisy_vs_noiseless}~(a)).
With the \ac{PM} model, using only $N_{\rm b} = 32$ bath sites already gives a good approximation on a large time range (with a little loss of accuracy at short times).
In the frequency domain (Fig.~\ref{fig:noisy_vs_noiseless}~(b)), the short-time limitation translates into a poor energy resolution, as evidenced by
the Fourier series of $G^>(t)$ for $N_{\rm b}=50$ closed bath sites, limited to the time range $[-25, 25]$.
By contrast, with the \ac{PM} model, the spectrum converges for $N_{\rm b} \le 32$.

The second advantage of open baths is the automatic state preparation.
In order to get correlation functions, one should prepare the relevant initial state before time evolution.
With closed baths, the ground (or Gibbs) state needs to be prepared, which usually requires deep circuits and advanced optimization methods (see e.g \cite{Besserve2024}).
By contrast, with open baths, the relevant state is the steady state, which is automatically prepared given a long enough relaxation $t_{\rm prep}$ before the actual \ac{GF} computation.
It turns out that the convergence to this steady state is exponential: 
the error made on the \ac{GF} for a finite $t_{\rm prep}$ vanishes exponentially with a rate $\Gamma \approx \gamma^2 / \pi \approx 0.11$ that varies only slightly with the number $N_{\rm b}$ of bath sites (Fig.~\ref{fig:noisy_vs_noiseless}~(c);
see Apps.~\ref{app:relaxation_rate}-\ref{app:error-def} for details and definitions;
emitter qubits are initialized to $\ket{1}$, absorber and impurity qubits to $\ket{0}$).

\begin{figure}[t]
    \centering
     \includegraphics[width=\linewidth]{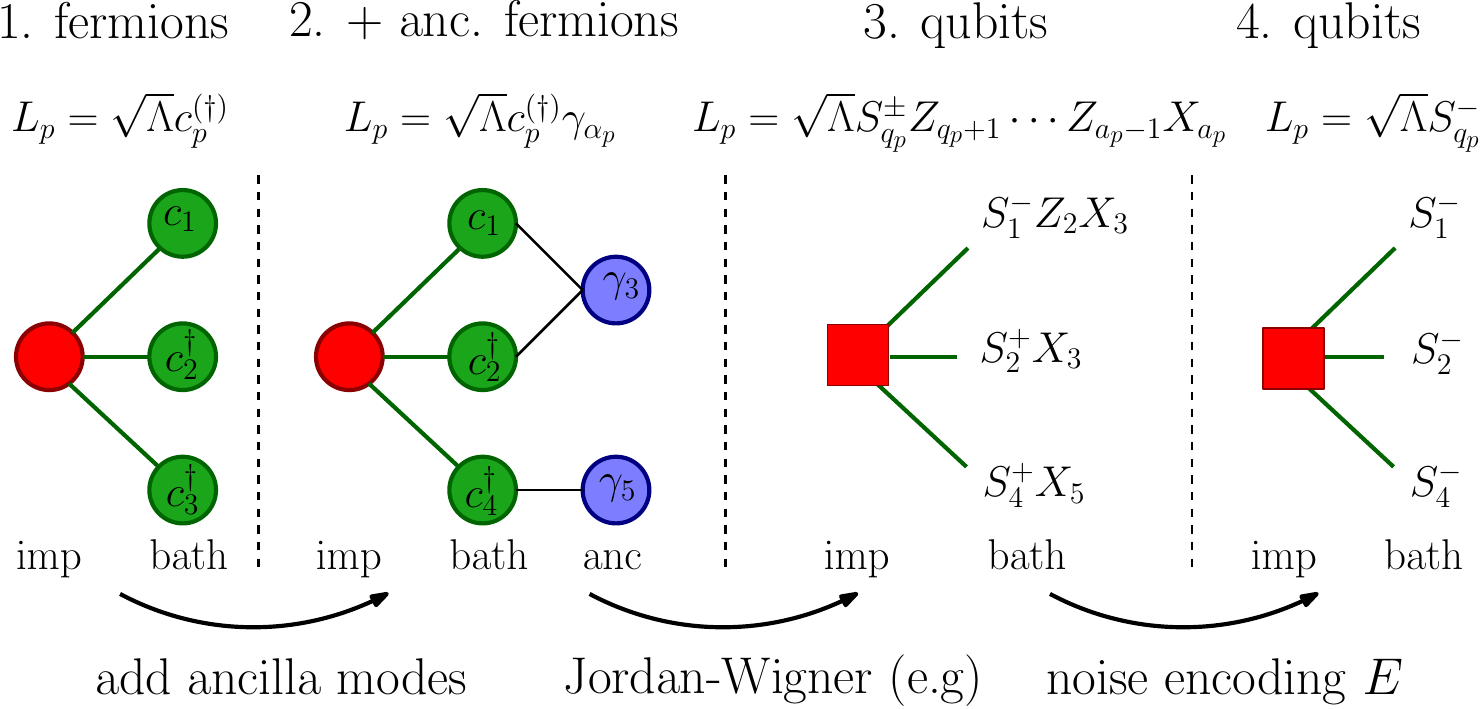}
    \caption{
        \label{fig:encoding}
        \textit{From fermion dissipation to qubit noise.} Three steps to map jump operators $c^{(\dag)}$ onto amplitude damping $S^-$: addition of ancilla fermions (Eq. \eqref{eq:anc_fermions}) to make jump operators local; fermion-qubit encoding; mapping onto $S^-$ by noise encoding $E$.
	  }
\end{figure}

\section{Noise harvesting quantum circuit}
To simulate the unitary part of the target Lindblad dynamics (Eq.~\eqref{eq:Lindblad}) with a quantum processor, we follow a standard Trotter approach.
As for the dissipative part, embodied by the jump operators $L\propto c^\dagger$ or $\propto c$, we would like to use the processor's natural noise to reproduce it.
Amplitude damping, defined by qubit jump operators $L\propto S^- \equiv \ket{0}\bra{1}$, is the most straightforward noise type that we can aim for:
Indeed, through (e.g) a \acf{JW} transformation, $c$ and $c^\dag$ are mapped onto operators $S^{\pm} Z Z \cdots Z$.
To map them onto the sought-after $S^-$, we first need to remove the \ac{JW} strings of $Z$ operators and possibly change $S^{+}$ to $S^{-}$.
Second, we need to tune the coefficient in front of the jump operator so that the qubit noise magnitude matches the targeted dissipation rate.

\subsection{Turning $c^{(\dagger)}$ into $S^-$}
We address the first challenge by performing three successive transformations, illustrated in Fig.~\ref{fig:encoding}, leaving the dynamics of the \ac{PM} model unchanged.
First, in order to make jump operators local, we introduce ancilla fermionic modes (step $1 \rightarrow 2$).
Each bath mode $p$ is associated with an ancilla mode of index $\alpha_p$, which can be shared between several bath modes.
Jump operators are transformed into
\begin{equation}
    \label{eq:anc_fermions}
    L_p \rightarrow L_p \gamma_{\alpha_p},
\end{equation}
with $\gamma_{\alpha_p}$ denoting Majorana operators acting on ancilla mode $\alpha_p$.
Because these are now two-fermion operators, they can be mapped (e.g via \ac{JW}) onto bounded-support qubit operators.
These ancillas do not affect the model dynamics, as shown in App.~\ref{app:proof_ancillas}.
They can be thought of as buffer modes between the bath and its environment, bookkeeping changes in fermion parity.
Ancilla modes will be represented by noiseless ancilla qubits. 
The number and distribution of ancilla modes is up to choice, but to make sure that jump operators are $\order{K}$-local, we make $\lceil N_{\rm b} / K \rceil = N_{\rm anc} + 1$ blocks of $K$ consecutive bath modes and associate them with an ancilla mode, placed at the end of the group.
Adding many ancillas makes jump operators very local, reducing the final number of gates, but requires more noiseless qubits.
The last ancilla, being at the end of the fermion order, brings no advantage and is discarded, so we have $N_{\rm anc}$ ancillas. 

Second, we apply e.g. a \ac{JW} transformation (step $2 \rightarrow 3$ in Fig.~\ref{fig:encoding}).
The Hamiltonian becomes $H_{\rm JW}$ and the jump operators become
\begin{equation}
    \label{eq:jump_op_after_JW}
    S^\pm_{q_p} \otimes \qty(\bigotimes_{q_p < k < a_p}  Z_k ) \otimes X_{a_p},
\end{equation}
where $q_p$ and $a_p$ are the qubits to which fermionic modes $p$ and $\alpha_p$ are mapped.
We note that other fermion-qubit transformations, including the Bravyi-Kitaev~\cite{bravyi_fermionic_2002} and Verstraete-Cirac~\cite{verstraete_mapping_2005} transformations, would also lead to expressions of the form $S^\pm_{q_p} \bigotimes_i P_{n_i}$, 
where $P_{n_i}$ is a Pauli operator acting on qubit $n_i$ and $\{n_i\}$ a list of qubits that does not contain $q_p$. 
Such a form is required for the following step.

\begin{figure}[t]
    \includegraphics[width=\linewidth]{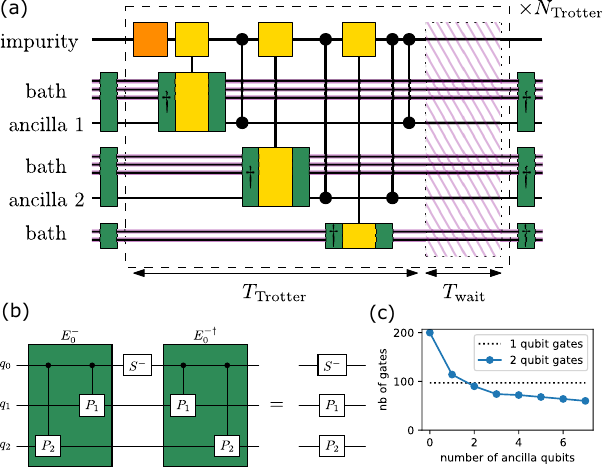}
    \caption{
        \label{fig:circuit-Trotter}
        \label{fig:noise-encoding}
        \textit{Quantum circuit to harvest qubit noise.}
        (a) Quantum circuit reproducing the dynamics of Eq.~\eqref{eq:Lindblad} with $N_\mathrm{b} =8$ bath sites.
        Noisy qubits (purple wires) are encoded (green boxes) to interpret  amplitude damping as fermion dissipation.
        The encoded Hamiltonian is Trotterized (dashed box), with impurity-only (orange box) and impurity-bath (yellow boxes) dynamics.
        (The CZ gates enforce the fermionic algebra between bath groups, see 
        App.~\ref{app:star_geometry_trick}).
        The dissipation rate is tuned by waiting after each Trotter step (zebra box).
        (b) Circuit version of Eq.~\eqref{eq:encoding_jump_op}, illustrated on 3 qubits ($P_1$ and $P_2$ are arbitrary Pauli operators).
        (c) Reduction of the number of 2-qubit gates per Trotter step with $N_\mathrm{anc}$, for $N_{\rm b} = 8$.
        }
\end{figure}

Third, we design a unitary transform, $E_p^\pm$ (that we dub "noise encoding"), that transforms the jump operator Eq.~\eqref{eq:jump_op_after_JW} into the amplitude damping jump operator (step $3 \rightarrow 4$ in Fig.~\ref{fig:encoding}):
\begin{equation}
    \label{eq:encoding_jump_op}
     (E_p^\pm)^\dag \; S^\pm_{q_p} \bigotimes_i P_{n_i} \; E_p^{\pm} = S_{q_p}^-.
\end{equation}
As any unitary transform, it acts like a change of basis and therefore does not affect the dynamics.
$E_p^-$ is realized by a circuit, illustrated in Fig.~\ref{fig:noise-encoding} (b),
which applies every $P_{n_i}$ controlled by the bath qubit $q_p$.
$E_p^+$ is identical, with a $X$ gate applied on the bath qubit beforehand.
The full noise-encoding operator is simply 
 $E = \prod_p E_p^{s_p}$,
where $s_p = +$ for emitter and $-$ for absorber bath sites.

We have thus mapped fermionic jump operators onto qubit amplitude damping jump operators. 
The noise encoding $E$ is applied to the Hamiltonian $H_\mathrm{JW} \rightarrow \tilde H = E^\dag  H_\mathrm{JW} E$ and the initial state $\ket{\phi} \rightarrow \tilde{\ket{\phi}} = E^\dag  \ket{\phi}$ to complete the transformation. 
Importantly, since $E$ is a product of $\order{K}$-local tranformations, it preserves the locality of $H_\mathrm{JW}$. 

\subsection{Harvesting noise with a waiting time}
We now solve the second challenge, i.e. matching the \ac{PM} model's dissipation rate $\Lambda$ with the hardware's amplitude damping magnitude, parameterized by its $T_1$ coherence time.
We achieve this by inserting waiting times $T_\mathrm{wait}$ (zebra area in Fig.~\ref{fig:circuit-Trotter} (a)) in our Trotter evolution to adjust the overall noise amplitude:
given the physical duration of a Trotter step  $T_{\rm Trotter}$, we tune $T_{\rm wait}$ so that $(T_{\rm Trotter} + T_{\rm wait})/T_1 = \Lambda \tau$,
where $\tau=t/N_\mathrm{trotter}$ is the "logical" duration of a Trotter step.
This allows us to mimic all dissipation rates $\Lambda$ above a minimal value $\Lambda_{\rm min} = T_{\rm Trotter} / T_1 \tau$.
This minimal value gives a limitation on the sharpness of the bath spectra that can be captured, as features on energy scales smaller than $\Lambda_{\rm min}$ will be smoothed out in the fitting procedure.
This minimal dissipation rate can be improved by increasing the Trotter step $\tau$, with the side effect of increasing Trotter error, or by increasing the qubit coherence time $T_1$, if experimentally possible, or by reducing the execution time (through e.g. compilation) of a Trotter step $T_{\rm Trotter}$.

The corresponding circuit thus reproduces the dynamics of Eq.~\eqref{eq:Lindblad}, with an encoding ($E$) circuit,
Trotter step circuits corresponding to the encoded Hamiltonian $\tilde H$, and  decoding ($E^\dag$) circuits.
An example is shown in Fig.~\ref{fig:circuit-Trotter} (a), using the \ac{JW} transformation, a single impurity site, $N_{\rm b} = 8$, $N_{\rm anc} = 2$ and $K=3$.
Noise occurring within an encoding-decoding pair is interpreted as dissipation, whereas noise occurring outside, e.g. during impurity-bath coupling circuits or encoding/decoding circuits, produces error.
$N_\mathrm{anc}$ can be adjusted: adding ancillas reduces the depth of the encoding circuit from quadratic in the bath size to linear, at fixed $K$.
In addition, for $K = \order{1}$, impurity-bath coupling terms have a support $K$ times smaller with ancillas, which translates into fewer 2-qubit gates, as shown in Fig.~\ref{fig:circuit-Trotter} (c).
This illustrates the advantage of ancillas with a moderate $K$, although it requires additional noiseless qubits
(see App.~\ref{app:encoding_complexity_with_ancillas}
for more details).
With this time-evolution circuit, the impurity \acp{GF} can be computed using e.g. a Hadamard test (details in
App.~\ref{app:hadamard_test}) 
or Krylov-based methods~\cite{endo2020, jamet2023}.

\subsection{Asymptotic advantage over closed bath simulation}
The ability of the \ac{PM} model to reach longer time dynamics with fewer bath sites
still holds in the presence of Trotterization and qubit noise errors.

In both methods, qubit noise causes a noise error proportional to the number of gates applied.
At fixed Trotter step, to reach a target time $t$, a closed bath implies a gate count that scales as $\propto t^2$, while our noise-harvesting method requires only a $\propto t$ scaling.
Indeed, in the former case, the bath size, whence the number of gates in a Trotter step, must grow linearly with $t$, and so does the number of Trotter steps to keep the Trotter error constant.
In the latter case, we can work at a fixed bath size, leading to a runtime (depth and waiting time) and a noise error that grow only linearly with $t$.

These scalings can be improved by optimizing the Trotter step and the bath size (see
App.~\ref{app:long-time-advantage}).
This leads to an improved $t^{3/2}$ scaling of the total error for the closed-bath method, but also an improved $t^{2/3}$ scaling for our noise-harvesting method.
Incidentally, these scalings imply that our method is advantageous both at fixed or optimized Trotter step and bath size.

\begin{figure}[t]
    \centering
    \includegraphics[width=\linewidth]{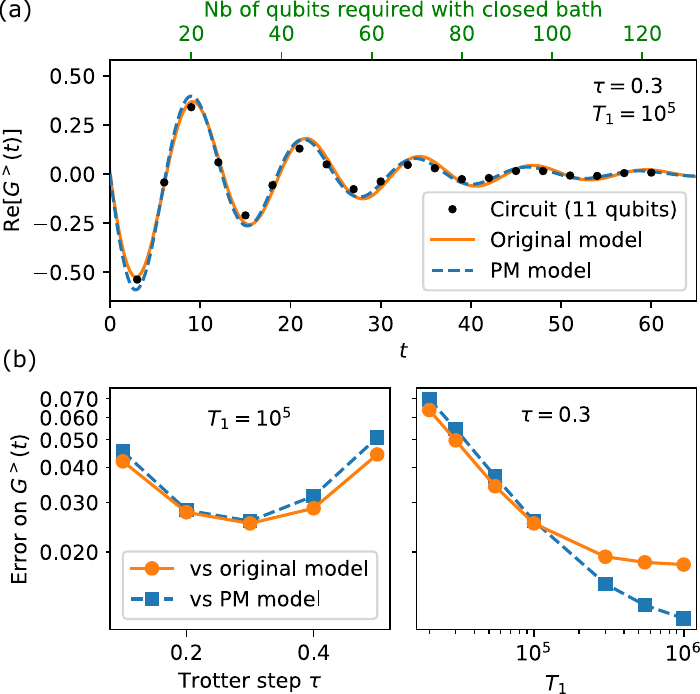}
    \caption{
        \label{fig:simu_quantum_circuit}
        \textit{Dynamics obtained through noise harvesting.}
        (a) \ac{GF} of the \ac{RLM} obtained with the circuit of Fig.~\ref{fig:circuit-Trotter}, with $N_{\rm b} = 8$,  $N_{\rm anc} = 1$ (dots), exact original (plain line) and \ac{PM} (dashed line) model dynamics.
        Top axis: number of qubits required with a closed bath.
        (b) Error vs Trotter step (left) or 
        vs coherence time $T_1$ (right).
	    }
\end{figure}

\section{Numerical illustration with a noisy circuit}
We numerically simulate our circuit to obtain the  greater \ac{GF} of the \ac{RLM} with amplitude damping after and in-between gates, with $T_1 = 10^5$ in units of the single-qubit gate duration.
We assume a 2-qubit gate duration $T_{\rm 2qb} = 10$, and use $N_{\rm b} = 8$ bath modes and $N_{\rm anc} = 1$ ancilla mode (thus halving the number of 2-qubit gates, see Fig.~\ref{fig:circuit-Trotter} (c)). 
The Trotter step is $\tau = 0.3$, and the preparation time $t_{\rm prep} = 30$ ensures negligible state preparation error.

We observe, in Fig.~\ref{fig:simu_quantum_circuit}~(a), a quantitatively close agreement between our noisy circuit and the original (before fit) and \ac{PM} (after fit) model dynamics.
This holds even in the long-time regime despite the very small (11) number of qubits.
A closed bath would require more than 120 qubits to obtain the same agreement (up to $t=60$), as evidenced in Fig.~\ref{fig:noisy_vs_noiseless}~(a).
We emphasize that the reduction in the number of qubits is proportional to the time one wants to reach.

The impact of the Trotter step $\tau$ and of the coherence time $T_1$ on the error (defined in App.~\ref{app:error-def}) compared to the original and \ac{PM} model dynamics is shown in Fig.~\ref{fig:simu_quantum_circuit}~(b).
As $\tau$ gets larger, the number of gates goes down, reducing the noise error (noise occurring during gate application), but the Trotter error increases.
This tradeoff leads to an optimal Trotter step $\tau_{\rm opt} \approx 0.3$ with our parameters.
At fixed $\tau$, the error follows a $1/T_1$ behavior (right panel), with saturation at large $T_1$. 
Indeed, increasing $T_1$ only changes the waiting time, but not the duration of the gates.
Therefore, the time intervals during which noise creates errors is unchanged, so that the noise error must decrease as $1/T_1$.
At large $T_1$, the noise error becomes negligible compared to the Trotter error and the error caused by the approximate fitting of the \acp{HF}, which explains the saturation of total error.
This saturation level depends whether fitting error is included (circles) or not (squares).
The difference therefore gives an estimate of the fitting error, here $\sim 10^{-2}$.

\section{Conclusion}
This work explores the possibility and potential advantages of harvesting qubit noise to produce controlled fermionic dissipation, an important ingredient of many condensed-matter models.
We provide a noise-harvesting quantum algorithm solving impurity models, the bottleneck of \ac{DMFT}.
Since it reproduces the real-time dynamics, it could be applied to equilibrium as well as non-equilibrium problems.

We demonstrate three advantages:
Reducing (by an order of magnitude in our test-case) the number of qubits,
unlocking access to long time dynamics, and
replacing ground-state preparation by the natural relaxation to a steady state.
Even though we studied a simple case, these advantages also hold with more structured baths that are relevant to \ac{DMFT}.
Indeed, the revival effect with closed baths will always lead to unbounded number of qubits in the long time dynamics.
Also, preparation of the relevant state by relaxation is guaranteed by the very definition of the problem to solve: a system that reached an equilibrium with a bath.

An experimental realization could make use of a hybrid architecture, where some good-quality qubits coexist with many lower-quality, $T_1$-limited (i.e. free of pure dephasing) qubits~\cite{kakuyanagi_dephasing_2007, kubica_erasure_2023}.
Due to its ability to tune qubit quality, superconducting technologies seem the most suitable.
With non-$T_1$-limited qubits, mitigation of pure dephasing becomes necessary, and can be performed with e.g. zero-noise extrapolation~\cite{temme_error_2017}.
Alternatively, a partial quantum error correction code~\cite{bultrini_battle_2023, koukoulekidis_framework_2023}, with a blend of noisy and clean logical qubits, and with a bias toward some noise channels could provide a way to perform this on standard hardware.
Furthermore, the robustness of our method to more realistic noise models, with second order processes, crosstalk, leakage and measurement errors, should be studied.


\acknowledgments
We acknowledge useful discussions with O. Parcollet and D. Est\`eve.
This work is part of HQI initiative (www.hqi.fr) and is supported by France 2030 under the French National Research Agency award number ANR-22-PNCQ0002.
The quantum circuit emulations were performed on the Eviden Qaptiva platform.

\appendix 

\section{Theoretical background}
\label{app:theory}
\acresetall

This technical appendix presents derivations of known results necessary for our purposes.

\subsection{Fermionic Green's function in open quantum systems with Markovian environment}
\label{app:def_gf_open_system}

In this section we recall the explicit expression of fermionic Green's functions in open quantum systems coupled to a Markovian environment.
A similar derivation for bosons or spins can be found in standard textbooks~\cite{breuerTheoryOpenQuantum2007}.

Correlation functions of an open quantum system $\systS$ can be properly defined as the corresponding correlation functions in a dilation $\systS \cup \systE$, where $\systE$ can be thought of as an environment, which follows unitary dynamics.
If the reduced dynamics on $\systS$ is Markovian, the \ac{QRT} provides an explicit formula for such correlation functions using only the reduced density matrix $\rho(t)$~\cite{breuerTheoryOpenQuantum2007}.

This result extends to fermionic systems with a few adaptations, caused by the nonlocal nature of fermionic operators.
Denote $\mathcal{L}$ the Lindbladian of $\systS$, and define $\altlindblad[\bullet] = (-1)^{N_\systS} \mathcal{L} \qty[(-1)^{N_\systS} \bullet]$, with $N_{\systS}$ the operator giving the number of fermions in $\systS$.
Consider two anticommuting fermionic operators $A$ and $B$ acting on $\systS$.
Their extension to $\systS \cup \systE$ is denoted $\tilde A$ and $\tilde B$.
According to the fermionic \ac{QRT}~\cite{schwarz_lindblad-driven_2016} (see Sec.~\ref{app:qrt}), the correlation function between $A$ and $B$ is
\begin{equation}
\label{eq:gf_lindblad_expr}
\begin{split}
    &\expval{\tilde A(t) \tilde B(t')}_{\systS \cup \systE} 
    \\
    &\quad\quad=
    \begin{cases}
        \Tr\qty(A e^{\altlindblad (t - t')} \qty[B e^{\mathcal{L} t'} [\rho_0]]) &\qq{if} t \ge t'
        \\
        \Tr\qty(B e^{\altlindblad (t' - t)} \qty[A^\dag e^{\mathcal{L} t} [\rho_0]^\dag]^\dag) &\qq{if} t < t'.
    \end{cases}
\end{split}
\end{equation}
for times $t, t' \ge 0$.
$\rho_0$ is the density matrix of $\systS$ at time $t=0$.
For commuting fermionic operators $A$ and $B$, $\altlindblad$ is replaced by $\mathcal{L}$, as would be obtained with the standard \ac{QRT}.

Without loss of generality, the Lindbladian can be written
\begin{gather}
    \label{eq:lindblad_app}
    \mathcal{L}[\bullet] = -i \comm{H}{\bullet} + \sum_k \qty[2 L_k^\nodag \bullet L_k^\dag - \qty{L_k^\dag L_k^\nodag , \bullet}]
\end{gather}
with jump operators $L_k$, and a Hamiltonian $H$.
Due to the parity superselection rule, $H$ must conserve the parity $(-1)^{N_\systS}$.
If all jump operators $L_k$ also conserve this parity, then we simply have $\altlindblad = \mathcal{L}$.
If, on the contrary, they all anticommute with this parity, then we have
\begin{equation}
    \label{eq:alt_lindblad}
    \altlindblad[\bullet] = -i \comm{H}{\bullet} - \sum_k \qty[ 2 L_k^\nodag \bullet L_k^\dag + \{L_k^\dag L_k^\nodag, \bullet\}],
\end{equation}
which is the same as Eq.~\eqref{eq:lindblad_app} with a different sign for the second term.

\subsection{Fermionic quantum regression theorem}
\label{app:qrt}

For completeness, we give a proof of the fermionic \ac{QRT}, i.e. of Eq.~\eqref{eq:gf_lindblad_expr}. It can also be found in Ref.~\cite[Appendix B]{schwarz_lindblad-driven_2016}.

Assume the unitary time evolution between times $t'$ and $t$ in the dilated space $\systS \cup \systE$ is represented by the operator $U_{t - t'}$.
The \ac{QRT} states that, under the same assumptions that makes $\systE$ Markovian, the reduced dynamics of any operator acting on $\systS$, i.e. $\tilde C = C_\systS \otimes \identity_\systE$, is given by the same Lindbladian $\mathcal{L}$ that describes the evolution of the reduced state.
More precisely, it states that
\begin{equation}
    \Tr_\systE \qty[U_{t - t'}^\dag \tilde C U_{t - t'}^\nodag \tilde \rho(t')]
    = e^{\mathcal{L} t} \qty[C_\systS \rho(t')]
\end{equation}
where $\tilde \rho$ is the density matrix of the full space $\systS\cup\systE$, and $\rho = \Tr_\systE[\tilde\rho]$ is the reduced density matrix on $\systS$. 

A consequence is that correlation functions between two operators acting on $\systS$, namely $\tilde A = A \otimes \identity_\systE$ and $\tilde B = B \otimes \identity_\systE$, are expressed in the reduced system as
\begin{align}
    &\expval{\tilde A(t) \tilde B(t')}_{\systS \cup \systE}
    = \Tr_{\systS \cup \systE} \qty[ \tilde A U_{t - t'}^\nodag \tilde B U_{t'}^\nodag \tilde \rho_0 U_{t}^\dag ]
    \\
    &\quad\quad=
    \begin{cases}
        \Tr_\systS\qty(A e^{\mathcal{L} (t - t')} \qty[B e^{\mathcal{L} t'} [\rho_0]]) &\qq{if} t \ge t'
        \\
        \Tr_\systS\qty(B e^{\mathcal{L} (t' - t)} \qty[e^{\mathcal{L} t} [\rho_0] A]) &\qq{if} t < t'.
    \end{cases}
\end{align}
with $t, t' \ge 0$, as can be shown using the cyclicity of the trace.

Nevertheless, this result cannot be used for fermionic \acp{GF}, due to the nonlocality of fermionic operators.
Let us take a convention for the order of fermion modes such that single-fermion operators acting on fermionic modes in $\systS$ are of the form $\tilde A = A \otimes (-1)^{N_\systE}$.
In these notations, $A$ is a single-fermion operator in the isolated system $\systS$ and $\tilde A$ is the corresponding operator acting in $\systS \cup \systE$.
$N_\systE$ is the number of fermion operators in $\systE$.
This form allows to respect anticommutation rules in $\systS \cup \systE$, and can be seen as deriving from the \ac{JW} transformation.
Then we get
\begin{equation}
\begin{split}
    &\expval{\tilde A(t) \tilde B(t')}_{\systS \cup \systE}
    \\
    &= \Tr_{\systS \cup \systE} \qty[ A \otimes (-1)^{N_\systE} U_{t - t'}^\nodag B \otimes (-1)^{N_\systE} U_{t'}^\nodag \tilde \rho_0 U_{t}^\dag ].
\end{split}
\end{equation}
Introducing $N_{\systS \cup \systE}$, the total number of fermion operators, and remembering the parity superselection rule, which imposes that the total fermion number parity commutes with $U$, we get
\begin{equation}
\begin{split}
    &\expval{\tilde A(t) \tilde B(t')}_{\systS \cup \systE}
    \\
    &= \Tr_{\systS \cup \systE} \qty[ A (-1)^{N_\systS} \otimes \identity_\systE  U_{t - t'}^\nodag (-1)^{N_\systS} B \otimes \identity_\systE  U_{t'}^\nodag \tilde \rho_0 U_{t}^\dag ].
\end{split}
\end{equation}
This form now respects the conditions to apply the \ac{QRT}, with operators $A (-1)^{N_\systS}$ and $(-1)^{N_\systS} B$.
Incorporating the $(-1)^{N_\systS}$ into an effective superoperator $\altlindblad[\bullet] = (-1)^{N_\systS} \mathcal{L} \qty[(-1)^{N_\systS} \bullet]$, we get Eq.~\eqref{eq:gf_lindblad_expr}.
To derive the expression for $t < t'$, we used the fact that $\mathcal{L}[\bullet]^\dag = \mathcal{L}[\bullet^\dag]$ and that $\mathcal{L}[\bullet (-1)^{N_\systS}] (-1)^{N_\systS} = \altlindblad[\bullet^\dag]^\dag$, and the cyclicity of the trace.

This derivation extends naturally to time-dependent Lindbladians.

\subsection{Validity of bath switching}
\label{app:wick}

The \ac{PM} representation relies on the important idea that, in some conditions, a bath can be replaced with a different (e.g. more compact) one without affecting the system's dynamics.
This was shown for bosonic ~\cite{tamascelli_nonperturbative_2018} and fermionic~\cite{chen_markovian_2019}  baths.
Here we aim at summing up the argument for the fermionic case in a pedagogical way.

Consider a system $\systS$ coupled to a bath $\systB$ with a linear coupling Hamiltonian
\begin{equation}
    H_{\systS\systE} = \sum_k A_k B_k
\end{equation}
where $A_k$ act on $\systS$ and $B_k$ act on $\systB$, and can be bosonic, fermionic or spin operators.
Assume the initial state is a product state between $\systS$ and $\systB$.
Then, perturbation theory on the coupling Hamiltonian tells us that the dynamics of $\systS$ depends on $\systB$ only through multi-point correlation functions\footnote{This includes single-point correlation functions, which are just the expectation values of the $B_k$ at every time.} of $\systB$ involving operators $B_k$, when $\systB$ is disconnected from $\systS$~\cite{rammer_quantum_2007}.
$\systB$ can thus be replaced with any other system $\systB'$ as long as all these correlation functions are the same.

Because this amounts to comparing an infinite number of correlation functions, this condition is impossible to verify if there is no structure to reduce the problem.
However, if $\systB$ and $\systB'$ are made of free particles, and if $B_k$ are linear in the creation or annihilation operators, Wick's theorem~\cite{rammer_quantum_2007} can be applied to write any multi-point correlation function as a sum of products of 2-point correlation functions.
In that case, it suffices that the 2-point correlation functions of $\systB$ and $\systB'$ involving $B_k$, which are called \acfp{HF}, and the expectation values of the $B_k$, match.
For fermions, expectation values of creation and annihilation operators are always zero, so we need only match \acp{HF}.

When $\systB$ and $\systB'$ are closed systems, Wick's theorem applies under the condition that their initial states be Gaussian and their Hamiltonians quadratic.
Nevertheless, the result above extends to some open systems, in particular those under Markovian dynamics with jump operators that are linear in creation and annihilation operators.
Indeed, such systems can be dilated into a larger closed system with unitary dynamics described by a quadratic Hamiltonian, for which Wick's theorem applies.
A proof for bosons can be found in Ref.~\cite{tamascelli_nonperturbative_2018}, and we give another version of this proof for fermions in the next section (Sec.~\ref{app:proof_dilation}).
This shows that Wick's theorem applies to such open quantum systems, with correlation functions defined as in Sec.~\ref{app:def_gf_open_system}.
Other kinds of open systems in which these results apply exist, e.g. systems under a quasi-Lindblad dynamics introduced in Ref.~\cite{park_quasi-lindblad_2024}.

\subsection{Dilation of Markovian system into a larger unitary system}
\label{app:proof_dilation}

We now give a proof that an open quantum system $\systS$ following Markovian dynamics can be dilated into a larger closed system following the Schr\"odinger equation.
It will then be clear that if $\systS$ is quadratic in (fermionic or bosonic) creation and annihilation operators, and if jump operators are linear in the same creation and annihilation operators, then the dilation is also quadratic, and can thus be subject to Wick's theorem.
A proof for bosons can be found in Ref.~\cite{tamascelli_nonperturbative_2018}.

Consider the generic Lindbladian of Eq.~\eqref{eq:lindblad_app}, assuming $N$ jump operators $L_1, \ldots, L_N$. The dynamics it generates can be approached arbitrarily close by a unitary dynamics with the following time-dependent Hamiltonian:
\begin{gather}
    \label{eq:ham_unitary_dyn}
    H_{\rm tot}(t) = H_0 + H_{\rm c}(t)
    \\
    H_0 = H + \sum_{k=1}^N \sum_{l \ge 0} \omega_{kl} b^\dag_{kl} b^\nodag_{kl}
    \\
    H_{\rm c}(t) = i \sqrt{2}\sum_{k=1}^N \sum_{l \ge 0} v_{l}(t) \qty[b^\dag_{kl} L_k^\nodag - L_k^\dag b_{kl}^\nodag]
\end{gather}
where $b^\dag_{kl}$ and $b^\nodag_{kl}$ are the creation and annihilation operators of an infinite number of additional bosonic or fermionic modes of energies $\omega_{kl}$, and $v_{l}(t)$ is a time dependent coupling to these modes.
These extra modes form an environment $\systE$, and we call them environment modes.
The environment modes are all initially in their vacuum state.
Time is divided in steps of duration $\tau$. Each jump operator $L_k$ is coupled to a single environment mode at a time, which changes at every time step.
Specifically, $v_l(t) = 1 / \sqrt{\tau}$ when $t$ is in the $l$th time step, and $v_l(t) = 0$ otherwise.
At time step $l$, we denote the environment modes currently coupled to the system as $\systE_l$.

If the system $\systS$ is made of fermionic modes and the jump operators are (anticommuting) fermionic operators (e.g. linear in creation and annihilation operators), then we must take the environment modes to be fermionic so that $H_{\rm c}(t)$ respects the parity superselection rule.
In any other situation, the environment modes are chosen to be bosonic.

We now prove that this dynamics, parameterized by $\tau$, approaches Eq.~\eqref{eq:lindblad_app} as $\tau \rightarrow 0$, after tracing out the environment modes.
Let us call $t$ the starting time of a given step, indexed $l$.
If $\psi(t)$ is the density matrix of the full system at that time, then the reduced state at the end of the time step is:
\begin{equation}
    \rho(t + \tau) = \Tr_{\systE} \qty(e^{-i H_{\rm tot}(t^+) \tau} \psi(t) e^{i H_{\rm tot}(t^+) \tau})
\end{equation}
where $H_{\rm tot}(t^+)$ is the Hamiltonian during step $l$.
The coupling term is $H_c(t^+) = i\sqrt{\frac{2}{\tau}} \sum_{k=1}^N \qty[b^\dag_{kl} L_k^\nodag - L_k^\dag b_{kl}^\nodag]$ and is proportional to $1/\sqrt{\tau}$.
$H_{\rm tot}(t^+)$ thus couples only $\systE_l$ to $\systS$, keeping the rest of $\systE$ isolated.
In that way, the trace over the rest of $\systE$ is trivial, and we get:
\begin{equation}
    \rho(t + \tau) = \Tr_{\systE_l} \qty(e^{-i H_{\rm tot}(t^+) \tau} \rho(t) \otimes \sigma_0 e^{i H_{\rm tot}(t^+) \tau})
\end{equation}
where $\sigma_0$ is the vacuum state of $\systE_l$, which has not changed before the beginning of step $l$.

Developing the exponentials leads to terms in integer and half-integer powers of $\tau$.
Terms with half-integer powers of $\tau$ must contain an odd number of $H_c(t^+)$, which causes them to vanish, as we now explain.
This odd number is split between those applied to the right-hand-side of $\psi(t)$ and those applied to the left-hand-side of $\psi(t)$.
These two groups necessarily differ in the number of $H_c(t^+)$ applied: one must be even and the other odd, so that the sum of the two is odd.
But $H_c(t^+)$ adds or removes one particle to $\systE_l$, so that, when tracing it out, $\sigma_0$ is sandwiched between states of different particle numbers, making the term vanish.

We are left with the integer powers of $\tau$, from which we are only interested in zeroth and first orders.
There are three contributions to the first order term (remember $H_c(t^+)$ contributes $\tau^{-1/2}$):
\begin{equation}
\begin{split}
    \rho(t + \tau) = \rho(t) & -i \tau \Tr_{\systE_l} \qty(\comm{H_0}{\rho(t) \otimes \sigma_0})
    \\
    &+ \tau^2  \Tr_{\systE_l} \qty(H_{\rm c}(t^+) \rho(t) \otimes \sigma_0 H_{\rm c}(t^+))
    \\
    &- \frac{\tau^2}{2}  \Tr_{\systE_l} \qty(\qty{H_{\rm c}(t^+)^2, \rho(t) \otimes \sigma_0})
    \\
    &+ \order{\tau^2}
\end{split}
\end{equation}
We expand $H_c(t^+)$ and keep only terms that do not vanish due to the projection onto the vacuum state $\sigma_0$.
We get, with the short-hand notation $\Delta\rho(t) = \rho(t + \tau) - \rho(t)$:
\begin{equation}
\begin{split}
    \frac{\Delta\rho(t)}{\tau} = & -i \comm{H}{\rho(t)}
    \\
    &+ 2\sum_k \Tr_{\systE_l} \qty(b^\dag_{kl} L_k^\nodag \rho(t) \otimes \sigma_0 L_k^\dag b_{kl}^\nodag)
    \\
    &- \sum_k \Tr_{\systE_l} \qty( \qty{L_k^\dag b_{kl}^\nodag b^\dag_{kl} L_k^\nodag, \rho(t) \otimes \sigma_0 } )
    \\
    &+ \order{\tau}
\end{split}
\end{equation}
Whether in the fermionic\footnote{This is true in spite of the non-local nature of fermionic operators. It can be shown with a similar representation of fermionic operaotrs as in App.~\ref{app:qrt}.} or the non-fermionic case, we can reduce the trace to $\systE_l$ operators only:
\begin{gather}
    \Tr_{\systE_l} \qty(b^\dag_{kl} L_k^\nodag \rho(t) \otimes \sigma_0 L_k^\dag b_{kl}^\nodag)
    = L_k^\nodag \rho(t) L_k^\dag \Tr_{\systE_l} \qty(b^\dag_{kl} \sigma_0 b_{kl}^\nodag)
    \\
    \begin{split}
    \Tr_{\systE_l} \qty( \qty{L_k^\dag b_{kl}^\nodag b^\dag_{kl} L_k^\nodag, \rho(t) \otimes \sigma_0 } )
    \\
    = \qty{L_k^\dag L_k^\nodag, \rho(t) } \Tr_{\systE_l} \qty(b_{kl}^\nodag b^\dag_{kl} \sigma_0)
    \end{split}
\end{gather}
such that we get
\begin{equation}
    \frac{\Delta\rho(t)}{\tau} = \mathcal{L}[\rho(t)] + \order{\tau}.
\end{equation}
This shows that the unitary dynamics is indeed arbitrarily close to the Lindblad dynamics, or more formally:
\begin{equation}
    \rho(t) = e^{\mathcal{L} t}[\rho(0)] + \order{\tau}.
\end{equation}
Note that this proof also holds for a time-dependent $H$, by considering the Trotterized Hamiltonian as a first-order approximation.

\subsection{Green's functions in pseudomode models}
\label{sec:diss_hybridization}

In this section we derive explicit expressions for Green's functions in generic fermionic \ac{PM} models.
These can be used to derive the form of \acp{HF} we use in our work, i.e. Eq.~\eqref{eq:diss_hyb_lesser_greater}.
Another derivation can also be found in Ref.~\cite{schwarz_lindblad-driven_2016}.

Consider a quadratic Hamiltonian
\begin{equation}
    H = \sum_{kl} h_{kl} c_k^\dag c_l^\nodag
\end{equation}
with $c_k^\nodag$ and $c_k^\dag$ fermionic annihilation and creation operators on mode $k$,
and a Lindbladian
\begin{equation}
\begin{split}
    \mathcal{L}[\rho] =& -i \comm{H}{\rho}
    + \sum_{kl} \Lambda^+_{kl} \qty[2 c_k^\dag \rho c_l^\nodag -\qty{c_l^\nodag c_k^\dag, \rho}]
    \\
    &+ \sum_{kl} \Lambda^-_{lk} \qty[2 c_k^\nodag \rho c_l^\dag -\qty{c_l^\dag c_k^\nodag, \rho}],
\end{split}
\end{equation}
where $\Lambda^\pm$ are two \ac{PSD} matrices of dissipation rates\footnote{Note how $\Lambda^-$ has inverted indices compared to $\Lambda^+$.}.
In a steady state, $\mathcal{L}[\rho] = 0$, so that the greater and lesser \acp{GF} are, according to Eq.~\eqref{eq:gf_lindblad_expr}, and for $t > 0$
\begin{gather}
    G^>_{ij}(t) = -i \Tr\qty(c_i^\nodag e^{\altlindblad t}\qty[c_j^\dag \rho])
    \\
    G^<_{ij}(t) = i \Tr\qty(c_i^\nodag e^{\altlindblad t}\qty[\rho c_j^\dag ])
\end{gather}
where $\altlindblad$ is given by Eq.~\eqref{eq:alt_lindblad}.

We define $\altlindblad^\dag$ the adjoint superoperator of $\altlindblad$ by the relation, valid for any operators $A$ and $B$, $\Tr(A \altlindblad[B]) = \Tr(\altlindblad^\dag[A] B)$.
It is used to rewrite the \acp{GF} in Heisenberg picture, introducing $c_i(t) = e^{\altlindblad^\dag t}[c_i]$,
\begin{gather}
    G^>_{ij}(t) = -i \Tr\qty(c_i^\nodag(t) c_j^\dag \rho)
    \\
    G^<_{ij}(t) = i \Tr\qty(c_j^\dag c_i^\nodag(t) \rho).
\end{gather}
Using the cyclicity of the trace and Eq.~\eqref{eq:alt_lindblad}, one can check that
\begin{equation}
\begin{split}
    \altlindblad^\dag[\bullet] ={}& i\comm{H}{\bullet} 
    - \sum_{kl} \Lambda^+_{kl} \qty[2 c_l^\nodag \bullet c_k^\dag +\qty{c_l^\nodag c_k^\dag, \bullet}]
    \\
    &- \sum_{kl} \Lambda^-_{lk} \qty[2 c_l^\dag \bullet c_k^\nodag + \qty{c_l^\dag c_k^\nodag, \bullet}].
\end{split}
\end{equation}
so that, after some fermionic algebra, we get
\begin{gather}
    \altlindblad^\dag[c_i] = \sum_l L_{il} c_l^\nodag,
    \\
    L_{kl} = -i h_{kl} - \qty(\Lambda_{kl}^+ + \Lambda_{kl}^-).
\end{gather}
Because $\Lambda^\pm$ are \ac{PSD}, the eigenvalues of $L$ must have real part $\le 0$.
From this we deduce that $c_i(t) = \sum_l \eval{e^{L t}}_{il} c_l(t=0)$, which yields matrix formulas for the lesser and greater \ac{GF} at $t > 0$,
\begin{gather}
    G^>(t) = -i e^{L t} (\identity - R),
    \\
    G^<(t) = i e^{L t} R,
    \\
    R_{kl} = \Tr\qty(c_l^\dag c_k^\nodag \rho),
\end{gather}
where $R$ is the \ac{1-RDM} of the steady state $\rho$.
Negative times can be obtained from the relations $G^{\gtrless}(-t) = -G^{\gtrless}(t)^\dag$.

Without knowledge of $R$, we can already derive the retarded \ac{GF}, defined as $G^R(t) = \theta(t) [G^>(t) - G^<(t)]$, with $\theta$ the Heaviside function, giving $G^R(t) = -i \theta(t) e^{Lt}$.
After Fourier transform, we get
\begin{equation}
    G^R(\omega) = \frac{1}{\omega - iL}.
\end{equation}
Similarly, we define the advanced \ac{GF} as $G^A(t) = G^R(-t)^\dag$ and its Fourier transform is
\begin{equation}
    G^A(\omega) = \frac{1}{\omega + iL^\dag}.
\end{equation}

To obtain the lesser and greater \ac{GF}, we need to characterize $R$, as it contains information about the steady state $\rho$.
Introducing $\mathcal{L}^\dag$, the adjoint superoperator of $\mathcal{L}$, we see that
\begin{equation}
\label{eq:steady_state_condition}
    \Tr\qty(\mathcal{L}^\dag\qty[c_l^\dag c_k^\nodag] \rho) = 0
\end{equation}
since $\mathcal{L}[\rho] = 0$.
Again, using the cyclicity of the trace, one can check that
\begin{equation}
\begin{split}
    \mathcal{L}^\dag[\bullet] ={}& i \comm{H}{\bullet}
    + \sum_{kl} \Lambda^+_{kl} \qty[2 c_l^\nodag \bullet c_k^\dag - \qty{c_l^\nodag c_k^\dag, \bullet}]
    \\
    &+ \sum_{kl} \Lambda^-_{lk} \qty[2 c_l^\dag \bullet c_k^\nodag - \qty{c_l^\dag c_k^\nodag, \bullet}]
\end{split}
\end{equation}
which leads to
\begin{equation}
    \mathcal{L}^\dag\qty[c_l^\dag c_k^\nodag] = \sum_n \qty(L_{nl}^\dag c_n^\dag c_k^\nodag + L_{kn} c_l^\dag c_n^\nodag) + 2 \Lambda_{kl}^+.
\end{equation}
Introducing this expression into Eq.~\eqref{eq:steady_state_condition}, we get a characterization for the steady state \ac{1-RDM},
\begin{equation}
\label{eq:steady_state_rdm_charact}
    R L^\dag + L R + 2 \Lambda^+ = 0.
\end{equation}

We now derive an expression for the lesser \ac{GF}.
Combining positive and negative times gives $G^<(t) = i\theta(t)e^{Lt}R +i\theta(-t) R e^{-L^\dag t} = -G^R(t) R + R G^A(t)$.
After Fourier transform, and factorizing $G^R(\omega)$ and $G^A(\omega)$, we get
\begin{equation}
    G^<(\omega) = G^R(\omega) \qty[(\omega - i L) R - R (\omega + i L^\dag)] G^A(\omega),
\end{equation}
which simplifies, using Eq.~\eqref{eq:steady_state_rdm_charact}, into
\begin{equation}
    G^<(\omega) = 2i G^R(\omega) \Lambda^+ G^A(\omega).
\end{equation}
Similarly, we find
\begin{equation}
    G^>(\omega) = -2i G^R(\omega) \Lambda^- G^A(\omega).
\end{equation}

In the case we are considering in the main text, this leads to an expression
\begin{equation}
\label{eq:diss_hyb_lesser_greater}
    \Delta_{\rm PM}^{\gtrless}(\omega) = \mp  2i\sum_{p\substack{\mathrm{even}\\\mathrm{odd}}} \frac{V_{p}^2 \Lambda} {|\omega - \epsilon_{p} + i \Lambda|^2}
\end{equation}
for the greater and lesser components of the \ac{HF}.

\section{Implementation details}
\acresetall

\subsection{Details of the bath fitting procedure}
\label{app:fitting}

\subsubsection{Closed bath}

To implement a thermal bath of spectral \ac{HF} $\Delta(\omega) = -\frac{1}{\pi}\Im[\Delta^R(\omega)]$ and inverse temperature $\beta$ with a closed bath, we discretize the frequency axis into energies $\epsilon_0, \ldots, \epsilon_{N_{\rm b}-1}$, and associate each $\epsilon_p$ to a bath mode $c_p$ coupled to the impurity by a term $V_p (c^\dag_p d + d^\dag c^\nodag_p)$, where
\begin{equation}
    V_p^2 = \int_{\epsilon_p'}^{\epsilon_{p+1}'} \dd{\omega} \Delta(\omega)
\end{equation}
and $\epsilon_0' = -D$, $\epsilon_{p-1} < \epsilon_p' < \epsilon_p$, $\epsilon_{N_{\rm b}}' = D$.
The distribution of energies $\{\epsilon_p\} \cup \{\epsilon_p'\}$ respects a centered density $f(\omega) \propto \theta(D + \omega) \theta(D - \omega) \exp(-2 \omega^2 / D^2)$, with $\theta$ the Heaviside function.
This ensures the relation
\begin{equation}
    \Delta(\omega) \approx \sum_{p=0}^{N_{\rm b}-1} V_p^2 \delta(\omega - \epsilon_p).
\end{equation}

The occupation of the bath is enforced by the choice of initial state.
For a thermal bath of inverse temperature $\beta$, the corresponding  Gibbs state of the impurity model, $\propto \exp(-\beta H_{\rm tot})$, with $H_{\rm tot}$ the hamiltonian of the impurity and closed bath, must be prepared.
For the \ac{RLM}, since it is quadratic, knowledge of the \ac{1-RDM} is enough to compute 1-particle correlation functions, and the 1-RDM can be obtained numerically by diagonalization.

\subsubsection{Open bath}

The representation of a bath with \acp{PM} boils down to the task of fitting each \ac{HF} $\Delta^<(\omega)$ and $\Delta^>(\omega)$ by a linear combination of Lorentzian functions. 
We add the constraint that all \acp{PM} should have the same dissipation rate $\Lambda$.

Eq.~\eqref{eq:diss_hyb_lesser_greater}
reveals that it is sufficient to fit the parameters of the emitters on the lesser \ac{HF}, and the parameters of the absorbers on the greater \ac{HF}.
In other words, the fit of $\Delta^<(\omega)$ and $\Delta^>(\omega)$ lead to separate \acp{PM}.
Thus, each fit may be done independently, and with the same procedure.
Note that in the \ac{RLM} we consider in this work, due to particle-hole symmetry, only one fit is required.
(The retarded \ac{HF} will follow automatically due to the relation $2i \Im{\Delta^R(\omega)} = \Delta^>(\omega) - \Delta^<(\omega)$ and the Kramers-Kroenig relations respected by $\Delta^R$.)

We denote $f(\omega)$ the (real positive) spectrum corresponding to $\Delta^<(\omega)$ or $\Delta^>(\omega)$, i.e. $f(\omega) = -i\Delta^<(\omega)$ or $f(\omega) = i\Delta^>(\omega)$.
In our fitting procedure, we discretize the frequency axis with a regular grid $\epsilon_0, \ldots, \epsilon_{N-1}$ which spans the frequency range on which $f(\omega) \ge \gamma^2 / 10$.
We take $N = N_{\rm b} / 2$.
This choice of frequency range allows to cut the exponential tail caused by the Fermi function.

Then, we search for the best parameters $v_p \ge 0$ and $\Lambda \ge 0$ that minimizes
\begin{equation}
    \chi^2(\vec{v}, \Lambda) = \int \dd{\omega} \qty[ f(\omega) - \sum_{p=0}^{N-1} \frac{v_p}{(\omega - \epsilon_p)^2 + \Lambda^2}]^2.
\end{equation}
At fixed $\Lambda$, we can perform the integral explicitly, giving
\begin{equation}
    \chi^2(\vec{v}, \Lambda) = \int \dd{\omega} f(\omega)^2
    - \sum_{p=0}^{N-1} v_p b_p 
    + \sum_{p,q=0}^{N-1} v_p Q_{pq} v_q,
\end{equation}
with
\begin{gather}
    b_p = 2 \int \dd{\omega} \frac{f(\omega)}{(\omega - \epsilon_p)^2 + \Lambda^2},
    \\
    Q_{pq} = \int \dd{\omega} \frac{1}{(\omega - \epsilon_p)^2 + \Lambda^2} \frac{1}{(\omega - \epsilon_q)^2 + \Lambda^2}.
\end{gather}
$Q$, as a Gram matrix, is \ac{PSD}, so that this is a convex optimization under convex constraint ($v_p \ge 0$) which is solved with CVXPY, a standard constrained quadratic programming library~\cite{boyd_convex_2004, diamond2016cvxpy}.
As convex optimization is stable, we can compute the optimal $v_p$ for arbitrary $\Lambda$, thus defining a new single-variable cost function $\chi^2(\Lambda)$, whose minimum we find by gradient descent.
This gives the optimal $v_p$ and $\Lambda$, from which we recover $V_p$ using the relation $v_p = 2V_p^2 \Lambda$.

\subsection{Relaxation rate}
\label{app:relaxation_rate}

\begin{figure}[t]
    \centering
    \includegraphics[width=\linewidth]{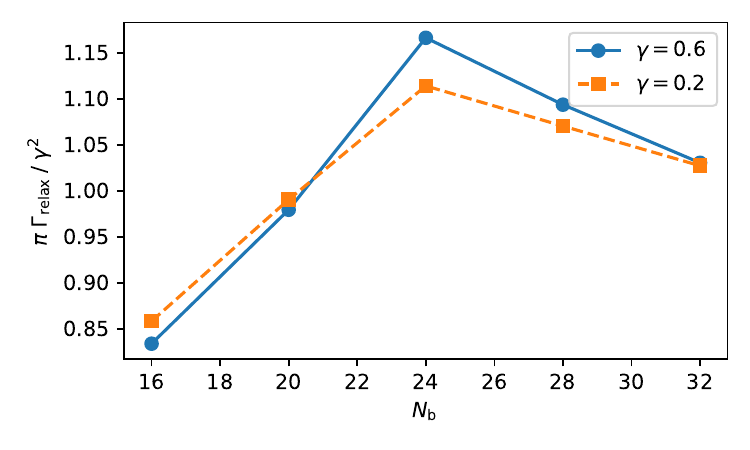}
    \caption{
        \label{fig:relaxation_rate}
        Relaxation rate in \ac{PM} models.
	    }
\end{figure}

The relaxation rate observed in Fig.~\ref{fig:noisy_vs_noiseless}~(c) varies slightly with the bath size $N_{\rm b}$, around the value $\gamma^2 / \pi$, as shown in Fig.~\ref{fig:relaxation_rate}.

Since it follows a Lindblad equation of the form Eq.~\eqref{eq:Lindblad}, the \ac{PM} model relaxes exponentially quickly toward a steady state.
For free fermions, characterized by a Hamiltonian $\sum_{ij} h_{ij} c^\dag_i c^\nodag_j$ and jump operators $\sqrt{\Lambda_i^+} c^\dag_i$ and $\sqrt{\Lambda_i^-} c^\nodag_i$, the relaxation rate (or rapidity) is lower bound by the lowest real part of the eigenvalues of the matrix~\cite{guo_solutions_2017}
\begin{equation}
    i h + \Lambda^+ + \Lambda^-
\end{equation}
with $\Lambda^{\pm}$ the diagonal matrix with diagonal entries $\Lambda_i^{\pm}$.
It can be shown that all its eigenvalues have positive real parts~\cite{guo_solutions_2017}, the lowest of which gives the slowest possible relaxation rate, $\Gamma$, depending on the initial state.

In Fig.~\ref{fig:noisy_vs_noiseless}~(c) and~\ref{fig:relaxation_rate} we used this $\Gamma$.
Figure~\ref{fig:noisy_vs_noiseless}~(c) confirms the exponential relaxation and shows it goes with the optimal rate.
We attribute variations of the relaxation rate to variations in the \ac{PM} \ac{HF}, due to a relatively small number of \acp{PM}.

\subsection{Definitions of errors}
\label{app:error-def}

The error shown in Fig.~\ref{fig:noisy_vs_noiseless}~(c) is defined as
\begin{equation}
    \sqrt{\int_0^{+\infty} \dd{t} \qty|G^>_{t_{\rm prep}}(t) - G^>_{t_{\rm relax=100}}(t)|^2}
\end{equation}
where $G^>_{t_{\rm prep}}(t)$ is the greater \ac{GF} between times $t + t_{\rm prep}$ and $t_{\rm prep}$.
Since in the \ac{RLM} under study the relaxation is exponential and the relaxation rate is $\Gamma \approx 0.11$ (see App.~\ref{app:relaxation_rate}), a relaxation time $t_{\rm prep}=100$ can be considered as infinite.

The error shown in Fig.~\ref{fig:simu_quantum_circuit}~(b) is defined as
\begin{equation}
    \varepsilon_{\rm tot} = \sqrt{\frac{1}{N_{\rm t}} \sum_{i=1}^{N_{\rm t}} \qty|G_{\rm circuit}^>(t_i) - G_{\rm exact}^>(t_i)|^2}
\end{equation}
$N_{\rm t} = 20$ is the number of time points $t_i$ used, which are those used in Fig.~\ref{fig:simu_quantum_circuit}~(a).
$G_{\rm circuit}^>$ is the greater \ac{GF} obtained by simulating our method, while $G_{\rm exact}^>$ is the exact greater \ac{GF} obtained numerically for either the original impurity model or the approximate \ac{PM} model.

\section{Proof that ancilla fermions do not disturb the system's dynamics}
\acresetall
\label{app:proof_ancillas}

We prove that the dynamics of the system (impurity and bath sites) is unaffected by the addition of ancilla fermions and their Majorana operators.

When introducing new modes, one has to be careful about how fermionic operators extend to the larger Hilbert space, in order to respect the anticommutation rules.
We denote by $\systS$ the system's Hilbert space, made of the impurity and bath modes, and $\systA$ the ancillas' Hilbert space.
For any single fermion operator $O_\systS$ acting on the system modes, we define its extension to the larger Hilbert space as $\tilde O_\systS = O_\systS \otimes \identity_\systA$.
Accordingly, for any single fermion operator $O_\systA$ acting on the ancilla modes, we define its extension to the larger Hilbert space as $\tilde O_\systA = (-1)^{N_\systS} \otimes O_\systA$, where $N_\systS$ is the number operator in the system alone.
This choice allows to respect the correct anticommutation relations in the full Hilbert space.

Consider the general case where a system of fermionic modes follows a Lindbladian dynamics with a Hamiltonian part $H$ and jump operators $L_p$ ($p=0, \ldots, N_{\rm b}-1$).
Adding ancilla modes, with indices $\{\alpha_p \qq{for} p=0, \ldots, N_{\rm b}-1 \}$ and with corresponding Majorana operators $\tilde \gamma_{\alpha_p}$, extends the Hilbert space such that the new hamiltonian is $\tilde H = H \otimes \identity_\systA$ and the new jump operators are $\tilde L_p \tilde \gamma_{\alpha_p} = (L_p \otimes \identity_\systA) \tilde \gamma_{\alpha_p} = L_p (-1)^{N_\systS} \otimes \gamma_{\alpha_p}$.
We denoted $\gamma_{\alpha_p}$ the Majorana operator restricted to the ancilla Hilbert space.
The density matrix of the system and ancillas $\tilde \rho$ then respects the Lindblad equation
\begin{equation}
    \dv{\tilde \rho}{t} = -i \comm{\tilde H}{\tilde \rho} + \sum_p \qty(2 \tilde L_p^\nodag \tilde \gamma_{\alpha_p} \tilde\rho \tilde \gamma_{\alpha_p} \tilde L_p^\dag  - \qty{\tilde L_p^\dag \tilde L_p^\nodag , \tilde \rho}),
\end{equation}
and we used the fact that $\tilde \gamma_{\alpha_p}^2 = \identity_{\systS \cup \systA}$ to cancel Majorana operators in the last term.
Tracing out the ancilla Hilbert space almost gives an equation for the density matrix of the system alone, $\rho$,
\begin{equation}
\begin{split}
    \dv{\rho}{t} =& -i \comm{H}{\rho} 
    \\
    &+ \sum_p \left(2 L_p (-1)^{N_\systS} \Tr_{\systA}\qty[\identity_\systS \otimes \gamma_{\alpha_p} \tilde \rho \identity_\systS \otimes \gamma_{\alpha_p}](-1)^{N_\systS} L_p^\dag  \right.
    \\
    &- \left. \qty{L_p^\dag L_p^\nodag , \rho} \right).
\end{split}
\end{equation}
To recover a Lindblad equation, the second term is simplified, first using the cyclicity of the trace to get $\Tr_{\systA}\qty[\identity_\systS \otimes \gamma_{\alpha_p} \tilde\rho \identity_\systS \otimes \gamma_{\alpha_p}] = \rho$, and then using the fermion superselection rule\footnote{There is no superposition of odd and even number of fermions.} in $\systS$, which states that $\rho$ commutes with $(-1)^{N_\systS}$.
We thus find that the reduced dynamics of the system is of the Lindblad type with Hamiltonian $H$ and jump operators $L_p$, as we wanted to show.

\section{Complexity of circuit with ancillas}
\acresetall
\label{app:encoding_complexity_with_ancillas}

We detail the consequences of using ancilla modes for circuit depth and locality, using \ac{JW}.
After \ac{JW}, the jump operator of bath site $p$ is given by Eq.~(7)
in the main text.
The corresponding encoding circuit $E_p^\pm$ is made of CZ gates between qubit $q_p$ and each qubit $k$ strictly between $q_p$ and $a_p$, and a CNOT gate applied on qubit $a_p$ with $q_p$ as the control qubit.
An $X$ gate follows on $q_p$ if the bath site is emitting.
Each encoding circuit $E_p^\pm$ therefore acts on at most $K+1$ qubits in the vicinity of $q_p$, whereas they were unbounded in absence of ancilla.
In addition, the full encoding circuit $E$ is made of $N_{\rm b} (K+1) / 2$ two-qubit gates, growing linearly with the bath size at fixed $K$, while it would be made of $N_{\rm b} (N_{\rm b} - 1) / 2$ two-qubit gates in the absence of ancilla.
This confirms the local nature of the encoding with ancilla, provided the number of ancilla modes is proportional to the number of bath modes, and placed regularly in the fermionic order.

The annihilation operator on bath mode $p$ is encoded into
\begin{equation}
    \tilde c_p = (-1)^{n^{\rm emit}_p + 1} S^-_{q_p} \qty(\bigotimes_{q_p < k < a_p} \!\!\! Z_k) \; X_{a_p} \qty(\bigotimes_{\substack{a \text{ ancilla} \\ a > a_p}} \!\!\! Z_a)
\end{equation}
where $n_p^{\rm emit}$ is the number of emitting bath sites strictly after bath site $p$, and the annihilation operator on the impurity mode $i$ is encoded into
\begin{equation}
    \tilde c_i = (-1)^{n^{\rm emit}_i} S_{q_i}^- \qty(\bigotimes_{q_i < k < n_{\rm imp}} \!\!\! Z_k) \qty(\bigotimes_{a \text{ ancilla}} \!\!\! Z_a)
\end{equation}
The coupling term between impurity mode $i$ and bath mode $p$ is then
\begin{align}
    \label{eq:encoded_coupling_with_ancillas}
    \tilde c_i^\dag \tilde c_p ={}& (-1)^{n^{\rm emit}_i} (-1)^{n^{\rm emit}_p + 1} 
    \tilde S_i^{{\rm imp}+}
    \qty(\bigotimes_{\substack{a \text{ ancilla} \\ a < a_p}} \!\!\! Z_a)
    \tilde S_p^{{\rm bath}-}
    \\
    \qq{with}& \tilde S_i^{{\rm imp}+} = S_{q_i}^+ \qty(\bigotimes_{q_i < k < n_{\rm imp}} \!\!\! Z_k)
    \\
    \qq{and}& \tilde S_p^{{\rm bath}-} = S^-_{q_p} \qty(\bigotimes_{q_p < k < a_p} \!\!\! Z_k) \; Z_{a_p} X_{a_p}
\end{align}
This term applies a chain of $Z$ operators on qubits between the impurity mode $i$ and the ancilla mode of bath $p$, but excluding every bath modes that are outside the block where $p$ lies.
In the worst case it acts on $\sim N_{\rm anc} + K = N_{\rm b} / K + K$ qubits.

Adding ancilla modes allows to reduce the overall depth of the circuit, by reducing the number of gates of the encoding circuit and by reducing the support of encoded hamiltonian terms.
The price to pay is twofold.
Not only does adding ancillas increase the total number of noiseless qubits required, but it also adds to the support of encoded hamiltonian terms and encoding circuit.
As a result, a minimal depth must be reached somewhere between a tightly local encoding that requires many ancillas (small $K$), and a more loosely local encoding with a few ancillas (large $K$).
Ultimately, available resources and hardware quality will dictate the best tradeoff between the number of noiseless qubits and depth.

\section{Circuit for impurities with star geometry}
\acresetall
\label{app:star_geometry_trick}

For impurities in the star geometry, as those we use within the \ac{PM} model, the circuits resulting from Trotterization have particular simplifications that allow to alleviate the effect of the chain of $Z$ operators on ancillas from Eq.~\eqref{eq:encoded_coupling_with_ancillas}.
We use notations introduced in Sec.~\ref{app:encoding_complexity_with_ancillas}.

For a Trotter step $\tau$, we need to successively apply impurity-bath coupling terms between impurity mode $i$ and bath mode $p$, with hopping energy $v_{ip}$, that read
\begin{equation}
    \label{eq:trotter_coupling}
    T_{i,p} = \exp(-iv_{ip} (\tilde c_i^\dag \tilde c_p + \text{h.c.}) \tau).
\end{equation}
Here, $\tilde c_i^\dag \tilde c_p$, given by Eq.~\eqref{eq:encoded_coupling_with_ancillas}, contains a chain of $Z$ operators acting on every ancilla mode before bath mode $p$.
Using the fact that
\begin{equation}
    \begin{quantikz}
        &\gate{S^\pm}& \qw{}
        \\
        &\gate{Z}& \qw{}
    \end{quantikz}
    =
    \begin{quantikz}
        &\ctrl{1} & \gate{S^\pm} & \ctrl{1}& \qw{}
        \\
        &\control{} & \ghost{Z}\qw{} & \control{}& \qw{}
    \end{quantikz},
\end{equation}
each $Z$ operator acting on ancilla $a$ can be replaced by a pair of \ac{CZ} gates between $i$ and $a$, surrounding the operator:
\begin{equation}
    \tilde c_i^\dag \tilde c_p = (-1)^{n^{\rm emit}_i} (-1)^{n^{\rm emit}_p + 1} 
    C_{i,p}
    \tilde S_i^{{\rm imp}+}
    \tilde S_p^{{\rm bath}-}
    C_{i,p}
\end{equation}
with
\begin{equation}
    C_{i,p} = \bigotimes_{\substack{a \text{ ancilla} \\ a < a_p}} \text{CZ}_{q_i, a},
\end{equation}
and where $\text{CZ}_{q_i, a}$ is the \ac{CZ} operator applied between qubits $q_i$ and $a$.
Note that \ac{CZ} gates always commute between each other.
$C_{i,p}$ is the product of \ac{CZ} operators between the impurity mode $i$ and every ancilla coming before the bath mode $p$.
Note that if $p$ and $p'$ are part of the same block, we have $a_p = a_{p'}$, such that $C_{i,p} = C_{i,p'}$.

These $C_{i,p}$ are unitary operators, so they can be pulled out of the exponential in Eq.~\eqref{eq:trotter_coupling}, i.e.
\begin{align}
    T_{i,p} ={}& C_{i,p} T_{i,p}' C_{i,p}
    \\
    T_{i,p}' ={}& \exp(-iv_{ip}' (\tilde S_i^{{\rm imp}+} \tilde S_p^{{\rm bath}-} + \text{h.c.}) \tau)
    \\
    v_{ip}' ={}& (-1)^{n^{\rm emit}_i} (-1)^{n^{\rm emit}_p + 1} v_{ip}
\end{align}
The advantage of this construction is that $T_{i,p}'$ requires no gate on ancilla qubits, whereas $T_{i,p}$ does.
In addition, the $C_{i,p}$ can be partially canceled out, as we explain below, so that the number of gates acting on ancillas is minimized.

For a single impurity mode $i=0$ and $N_{\rm b}$ bath modes $p=1, \ldots N_{\rm b}$, the full Trotter step is then the product
\begin{equation}
\begin{split}
    &T_{0,N_{\rm b}} \ldots T_{0,2} T_{0,1} =
    \\
    &C_{0,N_{\rm b}} T_{0,N_{\rm b}}' C_{0,N_{\rm b}}
    \ldots C_{0,2} T_{0,2}' C_{0,2} C_{0,1} T_{0,1}' C_{0,1}
\end{split}
\end{equation}
in which we can do the following simplifications.
If $p$ and $p+1$ are in the same block, $C_{i,p+1} = C_{i,p}$ so we have $C_{i,p+1} C_{i,p} = \identity$.
Otherwise, they are in neigbouring blocks, differing by only one \ac{CZ} gate, so that $C_{i,p+1} C_{i,p} = \text{CZ}_{q_i,a_p}$.
For example, ignoring bath qubits:
\begin{equation}
    \begin{quantikz}
        \lstick{imp} & \ctrl{1} \gategroup[3,steps=2,style={dashed,rounded
corners,fill=blue!20, inner
xsep=2pt},background]{$C_{i,p}$} & \ctrl{2}& \ctrl{1} \gategroup[4,steps=3,style={dashed,rounded
corners,fill=blue!20, inner
xsep=2pt},background]{$C_{i,p+1}$}&  \ctrl{2}&\ctrl{3}  & \qw{}
        \\
        \lstick{anc 1} & \control{}& \qw{} & \control{}& \qw{} & \qw{} & \qw{}
        \\
        \lstick{anc 2} & \qw{} & \control{}& \qw{} & \control{}& \qw{} & \qw{} 
        \\
        \lstick{anc 3} & \qw{} & \qw{} & \qw{} & \qw{} &\control{} & \qw{}
    \end{quantikz}
    =
    \begin{quantikz}
        &\ctrl{3} & \qw{}
        \\
        & \qw{}  & \qw{} 
        \\
        & \qw{}  & \qw{}
        \\
        &\control{} & \qw{}
    \end{quantikz}
\end{equation}
As a result, we only need one \ac{CZ} gate between each block, in addition to the gates from $C_{0, N_{\rm b}}$ at the end of the Trotter step, which lacks a partner to cancel with, giving a total of $2 N_{\rm anc}$ gates.
This construction is illustrated in the the circuit of Fig.~(4)
~(a) in the main text.
With more than one impurity site, this pattern can be repeated successively for each impurity site.

\section{The Hadamard test for computing fermionic Green's functions in open systems}
\acresetall
\label{app:hadamard_test}

In an open system submitted to a Markovian evolution superoperator $\mathcal{E}_t = e^{\mathcal{L}t}$, the correlation function between two anticommuting fermionic operators $A$ and $B$ is defined as (see Sec.~\ref{app:def_gf_open_system} and~\ref{app:qrt}):
\begin{equation}
\begin{split}
    &\expval{\tilde A(t) \tilde B(t')} =
    \\
    &\begin{cases}
        \Tr\qty(A (-1)^{N_\systS} \mathcal{E}_{t - t'} \qty[ (-1)^{N_\systS} B \mathcal{E}_{t'} [\rho_0]]) &\qq{if} t \ge t'
        \\
        \Tr\qty( (-1)^{N_\systS} B \mathcal{E}_{t' - t} \qty[ \mathcal{E}_t [\rho_0] A (-1)^{N_\systS}]) &\qq{if} t < t'.
    \end{cases}
\end{split}
\end{equation}
for two times $t, t' \ge 0$ and
with $N_{\systS}$ the operator giving the number of fermions in the system.
$\tilde A$ and $\tilde B$ are the operators extended to the system and environment.
We take the initial state to be pure $\rho_0 = \ket{\psi_0}\bra{\psi_0}$.

Assume we know a quantum circuit that produces the channel $\mathcal{E}_t$ for any time $t \ge 0$ (such as the one described in the main text), then we can compute this correlator using several Hadamard tests.
Specificaly, averaging measurement outcomes of the circuit
\begin{equation}
	\begin{quantikz}
		\lstick{$\ket{\psi_0}$} & \qwbundle{n} & \gate{\mathcal{E}_{t_1}} & \gate{U_1} & \gate{\mathcal{E}_{t_2}} & \gate{U_2^\dag} &   \qw{}  & \qw{}         
		\\
		\lstick{$\ket{0}$}      &     \qw{}    & \gate{H}                 & \ctrl{-1}  &   \gate{S^k}             & \octrl{-1}      & \gate{H} & \meter{}
	\end{quantikz}
\end{equation}
yields $\Re\qty[i^k\Tr\qty(U_2 \mathcal{E}_{t_2} \qty[U_1 \mathcal{E}_{t_1}[\rho_0]])]$,
where $U_1$ and $U_2$ are two unitaries and $k \in \{0, 1\}$.
Running it for both values of $k$ gives a way to compute $\Tr\qty(U_2 \mathcal{E}_{t_2} \qty[U_1 \mathcal{E}_{t_1}[\rho_0]])$.
By decomposing $\bar A = A (-1)^{N_\systS}$ and $\bar B = (-1)^{N_\systS} B$ into linear combinations of unitaries (it is always possible, e.g. in the Pauli basis), one can compute $\expval{\tilde A(t) \tilde B(t')}$ with such circuits.

We apply this for $A$ and $B$ being the impurity's creation or annihilation operators, after a \ac{JW} transformation.
We chose the convention of \ac{JW} such that $A$ and $B$ both contain a chain of $Z$ operators spanning the entire qubit register, i.e. we placed the impurity mode first in the fermion ordering and applied \ac{JW} with chains of $Z$ toward the end of the ordering.
This choice ensures that $\bar A$ and $\bar B$ are local to the impurity.
In the case we studied, with a single impurity mode, they are in fact single-qubit operators.

\section{Noise encoding}
\acresetall
\label{app:noise-encoding}

Here we discuss the physical interpretation of the noise encoding $E$ described in the main text.

Amplitude damping can be interpreted as random applications of the jump operator $S^- = \ket{0}\bra{1}$~\cite{Nielsen2010}.
Imagine this channel occurs between encoding and decoding, for an absorbing bath site.
The qubit register starts in a superposition $\alpha \ket{0}\ket{\phi_0} + \beta\ket{1}\ket{\phi_1}$, where $\ket{0}$ and $\ket{1}$ are states of the bath qubit, and $\ket{\phi_0}$ and $\ket{\phi_1}$ are states of the rest of the register, and $|\alpha|^2 + |\beta|^2 = 1$.
Encoding transforms $\ket{\phi_1} \rightarrow \bigotimes_i P_{n_i} \ket{\phi_1}$, but leaves the rest of the state unchanged.
The jump operator $S^-$ is applied with a probability $|\beta|^2$, projecting the state onto $\ket{0} \bigotimes_i P_{n_i} \ket{\phi_1}$, state on which the decoding gates have no effect.

Once in the state $\ket{0}$, additional $S^-$ operators cannot come (they happen with a zero probability). 
If no jump operator is applied, decoding gates simply undo the effect of the encoding gates.
Overall, the register sees a quantum channel with a jump operator $S^\pm_{q_p} \bigotimes_i P_{n_i}$, as we wanted.

\section{Advantage of using noise for long time dynamics}
\acresetall
\label{app:long-time-advantage}

Here we present a scaling analysis to evaluate the advantages and drawbacks of our noise-harvesting method when compared to a standard closed bath representation, on \ac{NISQ} hardware.
We look at the longest time dynamics that can be reached with an error budget $\varepsilon_{\rm max}$ and the resources (number of qubits and runtime) needed to reach it.
In both cases the dynamics is obtained by (first-order) Trotterization with a time step $\tau$, which can be optimized.
We do not expect higher-order Trotterization to change significantly our conclusions.
Qubits are assumed to be noisy, with an amplitude damping characteristic time $T_1$, and no pure dephasing, with the exception of impurity and ancilla qubits, which are noiseless (or quantum error corrected).

\subsection{Closed bath method}

With a closed bath representation, the number of bath sites $N_b$ must be proportional to the largest time $t$ one wants to reach, in order to avoid revivals~\cite{de_vega_how_2015}.
This means
\begin{equation}
\label{eq:Nb_t}
	N_{\rm b} = v t
\end{equation}
with $v$ some velocity.

There are two sources of error.
One is the (total) Trotter error, which for $N_{\rm Trotter}$ Trotter steps and a time step $\tau$ is
\begin{equation}
	\varepsilon_{\rm Trotter} = \alpha \tau^2 N_{\rm Trotter} = \alpha \tau t
\end{equation}
and $\alpha$ is a constant.
We used $t = N_{\rm Trotter} \tau$.
The other is the noise error, which is proportional to the runtime $T$ divided by the characteristic time of the noise $T_1$,
\begin{equation}
	\varepsilon_{\rm noise} = \beta\frac{T}{T_1}
\end{equation}
and $\beta$ is another constant.
The total error is
\begin{equation}
	\varepsilon = \varepsilon_{\rm Trotter} + \varepsilon_{\rm noise}.
\end{equation}
The physical time to apply one Trotter step is proportional to the number of bath sites $N_{\rm b}$, so that
\begin{equation}
	T = T_g N_{\rm Trotter} N_{\rm b} = v T_g \frac{t^2}{\tau}
\end{equation}
where $T_g$ is a constant which roughly corresponds to the time it takes to apply the coupling between the impurity and one bath site. 
The total error is then
\begin{equation}
	\varepsilon = \alpha \tau t + \frac{\beta v T_g}{T_1 \tau} t^2.
\end{equation}

One can see that for a given $\tau$, the noise error grows quadratically in $t$, due to the fact that not only more Trotter steps must be applied when increasing $t$, but more bath sites must be added too, leading to more gates in each Trotter step.
However, the Trotter error grows only linearly with $t$.
In fact, the balance between the two sources of errors can be tuned with the choice of $\tau$, so that a total error scaling as $t^{3/2}$ can be reached, as we now show.

Minimizing $\varepsilon$ as a function of $\tau$ leads (indeed, the extremum of a function $f(x) = a x + b/x$ is reached for $x_0$ such that $ax_0 = b / x_0$) to $\varepsilon_{\rm Trotter}(\tau_{\rm opt}) = \varepsilon_{\rm noise}(\tau_{\rm opt})$, from which we get
\begin{equation}
    \label{eq:H7}
    \tau_{\rm opt} = \sqrt{\frac{\beta v T_g t}{\alpha T_1}}.
\end{equation}
The error made is then
\begin{equation}
    \varepsilon_{\rm opt} = 2\varepsilon_{\rm Trotter} = 2 \sqrt{\frac{\alpha \beta v T_g t^3}{T_1}}
\end{equation}
The largest time dynamics is reached when the error budget is spent, which happens at a time
\begin{equation}
    t_{\rm max} = \qty(\frac{T_1 \varepsilon_{\rm max}^2}{4\alpha \beta v T_g})^{1/3}
\end{equation}
With the optimal choice of Trotter step, the runtime to reach $t$ is
\begin{equation}
    T = \sqrt{\frac{\alpha v T_g T_1 t^3}{\beta}}
\end{equation}

These scaling relations are summed up in the first column of Table~\ref{tab:summary_scalings}.

\begin{table}[t]
    \centering
    \begin{equation*}
    \begin{array}{c|c|c|c}
        & \text{closed bath} & \multicolumn{2}{c}{\text{noise-harvesting}}
        \\
        & & \text{fixed $N_{\rm b}$ and $\tau$} & \text{optimized $N_{\rm b}$ and $\tau$}
        \\
        \hline
        \varepsilon_{\rm opt}  & \sqrt{\dfrac{t^3}{T_1}} & t & \qty(\dfrac{t^2}{T_1})^{1/3}
        \\
        t_{\rm max} & \qty(T_1 \varepsilon_{\rm max}^2)^{1/3} & \varepsilon_{\rm max} \vphantom{b}^{(*)} & \sqrt{T_1 \varepsilon_{\rm max}^3}
        \\
        N_{\rm b} & t & \text{Const.} & \qty(\dfrac{T_1}{t^2})^{1/3}
        \\
        T & \sqrt{T_1 t^3} & t & (T_1^2 t^5)^{1/3}
    \end{array}
    \end{equation*}
    $\vphantom{b}^{(*)}$ assuming $\varepsilon_{\rm max} \gg \gamma / N_{\rm b}$.
    \caption{
        \label{tab:summary_scalings}
        Comparison of performance of the closed-bath approach versus our noise-harvesting approach to reach long time dynamics.
        All quantities are asymptotic behavior at large $t$ or small $\varepsilon_{\rm max}$.
    }
\end{table}

\subsection{Noise-harvesting method}

In our method, the \acp{HF} are fit with $N_{\rm b}$ Lorentzians of width $\Lambda$.
Assuming we need to cover the spectrum over an energy interval $D$, we need $D \approx \Lambda N_{\rm b}$.
Each Lorentzian corresponds to a bath site submitted to dissipation at a rate $\Lambda$.
This dissipation is obtained by the natural noise of qubits, which occurs with a characteristic time $T_1$.
Physical time is divided between application of gates, in which noise produces errors, and waiting time, in which noise produces the expected dissipation.
The number of noise events (e.g. jumps) occurring in each noisy qubit during the waiting time $T_{\rm wait}$ is $T_{\rm wait} / T_1$.
To get the correct dissipation rate, this number must equal the number of dissipative events occurring in the impurity dynamics during the Trotter time $\tau$, which is $\tau \Lambda$.
Therefore we have
\begin{equation}
	T_{\rm wait} = T_1 \tau \Lambda = \frac{T_1 D \tau}{N_{\rm b}}.
\end{equation}
The rest of the runtime is spent applying gates and is proportional to the number of bath sites.
The total runtime is then
\begin{align}
    T ={}& N_{\rm Trotter} \qty(T_g N_{\rm b} + T_{\rm wait})
    \\
	\label{eq:correct_noise}
    ={}& t \qty(\frac{N_{\rm b} T_g}{\tau} + \frac{T_1 D}{N_{\rm b}})
\end{align}
with $T_g$ the same constant as in the closed bath derivation.

There are three sources of error in this method.
The Trotter error is the same as in the previous section,
\begin{equation}
	\varepsilon_{\rm Trotter} = \alpha \tau t.
\end{equation}
The noise error is caused by noise events happening during gate application, and not in waiting time.
It is proportional to the total gate application time divided by $T_1$, and thus reads
\begin{equation}
    \label{eq:noise_error_2}
	\varepsilon_{\rm noise} = \beta \frac{t N_{\rm b} T_g}{\tau T_1}
\end{equation}
with $\beta$ the same constant as in the closed bath derivation.
The last source of error comes from the imperfect fit of the \acp{HF}.
It can be assumed to depends only on the granularity of the Lorentzians, so that
\begin{equation}
	\varepsilon_{\rm fit} = \gamma / N_{\rm b}
\end{equation}
with some constant $\gamma$.
The full error is thus
\begin{align}
    \varepsilon ={}& \varepsilon_{\rm Trotter} + \varepsilon_{\rm noise} + \varepsilon_{\rm fit}
     \\
    ={}& t \qty(\alpha \tau + \beta \frac{N_{\rm b} T_g}{\tau T_1}) + \frac{\gamma}{N_{\rm b}}
\end{align}

At fixed $\tau$ and $N_{\rm b}$, we see that the error is linear in $t$, which is better than in the closed bath method, even with optimized Trotter step.
The price to pay for this improved scaling is the constant fit error.
Also, the runtime $T$ grows only linearly with $t$, which again is better than the closed bath method where it is quadratic.
We conclude that for long time dynamics the noise-harvesting method brings an advantage both in runtime and number of qubits.
These scalings with fixed $N_{\rm b}$ and $\tau$ are summarized in the second column of Table~\ref{tab:summary_scalings}.

In the following we use the freedom in the choice of Trotter step and bath size to minimize error so as to reach the longest time dynamics possible within the error budget.
We will see that the error scaling can be improved to $t^{2/3}$ with a reduced bath size, if one is ready to give up on runtime.

\subsection{Optimizing for long time dynamics}

Minimizing $\varepsilon$ with respect to $\tau$ leads to $\varepsilon_{\rm Trotter} = \varepsilon_{\rm noise}$, and minimizing it with respect to $N_{\rm b}$ leads to $\varepsilon_{\rm noise} = \varepsilon_{\rm fit}$ (using the same property as used to obtain Eq.~\eqref{eq:H7}).
Therefore, the minimum is reached when the three errors are equal, giving the optimal values
\begin{gather}
    \tau_{\rm opt} = \qty(\frac{\gamma\beta T_g}{\alpha^2 T_1 t})^{1/3},
    \\
    \label{eq:opt_bath_size}
    N_{\rm b, opt} = \qty(\frac{\gamma^2 T_1}{\alpha \beta T_g t^2})^{1/3},
\end{gather}

Surprisingly, the optimal number of bath sites \emph{decreases} with $t$.
This is because a large bath is detrimental to the noise error, and even more so at large $t$ (see Eq.~\eqref{eq:noise_error_2}), whereas it improves the fit error, but independently of $t$.
At increasing $t$, the tradeoff is thus increasingly toward a small bath.

This does not mean the bath must be reduced to go to longer times.
Indeed, at fixed bath size and Trotter step, the error is a strictly increasing function of $t$.
This means one can just choose a fixed bath size $N_{\rm b, opt}(t = t_{\rm max})$ and Trotter step $\tau_{\rm opt}(t = t_{\rm max})$ to compute for all times within $[0, t_{\rm max}]$. We are then guaranteed the error is always within budget, even though it would not be minimal for $t < t_{\rm max}$.
This corresponds to choosing a bath size $N_{\rm b} = 3\gamma / \varepsilon_{\rm max}$, highlighting the fact that a smaller error budget requires a larger bath size.

With these values, the error is
\begin{equation}
    \varepsilon_{\rm opt} = 3 \varepsilon_{\rm Trotter} = 3 \qty(\frac{\alpha \beta \gamma T_g t^2}{T_1})^{1/3}
\end{equation}
which leads to a maximal time
\begin{equation}
    t_{\rm max} = \sqrt{\frac{T_1}{\alpha \beta \gamma T_g} \qty(\frac{\varepsilon_{\rm max}}{3})^3}.
\end{equation}

With the optimal bath size and Trotter step, the runtime to reach $t$ is
\begin{equation}
    T = \qty(\frac{\alpha \gamma T_g T_1^2}{\beta^2}t^2)^{1/3} \qty[1 + \frac{D \beta}{\gamma} t],
\end{equation}
the first term comes from gate applications and the second is waiting time.
At large $t$ the waiting time dominates.

The scalings with optimized $N_{\rm b}$ and $\tau$ for long time dynamics are summarized in the last column of Table~\ref{tab:summary_scalings}.
We see that the error scales as $t^{2/3}$, which is better than the closed bath scaling $t^{3/2}$, allowing to reach for a longer time dynamics with the same error budget (second row).
This improved scaling comes with a smaller number of qubits, scaling as $t^{-2/3}$ instead of $t$ in the closed bath case.
The price to pay to reach this very long time dynamics is runtime, which scales as $t^{5/3}$ compared to $t^{3/2}$ in the closed bath case.
However, note that the difference in power law is not large, namely $5/3 - 3/2 \approx 0.167$.

\bibliography{biblio}

\begin{thebibliography}{69}%
\makeatletter
\providecommand \@ifxundefined [1]{%
 \@ifx{#1\undefined}
}%
\providecommand \@ifnum [1]{%
 \ifnum #1\expandafter \@firstoftwo
 \else \expandafter \@secondoftwo
 \fi
}%
\providecommand \@ifx [1]{%
 \ifx #1\expandafter \@firstoftwo
 \else \expandafter \@secondoftwo
 \fi
}%
\providecommand \natexlab [1]{#1}%
\providecommand \enquote  [1]{``#1''}%
\providecommand \bibnamefont  [1]{#1}%
\providecommand \bibfnamefont [1]{#1}%
\providecommand \citenamefont [1]{#1}%
\providecommand \href@noop [0]{\@secondoftwo}%
\providecommand \href [0]{\begingroup \@sanitize@url \@href}%
\providecommand \@href[1]{\@@startlink{#1}\@@href}%
\providecommand \@@href[1]{\endgroup#1\@@endlink}%
\providecommand \@sanitize@url [0]{\catcode `\\12\catcode `\$12\catcode `\&12\catcode `\#12\catcode `\^12\catcode `\_12\catcode `\%12\relax}%
\providecommand \@@startlink[1]{}%
\providecommand \@@endlink[0]{}%
\providecommand \url  [0]{\begingroup\@sanitize@url \@url }%
\providecommand \@url [1]{\endgroup\@href {#1}{\urlprefix }}%
\providecommand \urlprefix  [0]{URL }%
\providecommand \Eprint [0]{\href }%
\providecommand \doibase [0]{https://doi.org/}%
\providecommand \selectlanguage [0]{\@gobble}%
\providecommand \bibinfo  [0]{\@secondoftwo}%
\providecommand \bibfield  [0]{\@secondoftwo}%
\providecommand \translation [1]{[#1]}%
\providecommand \BibitemOpen [0]{}%
\providecommand \bibitemStop [0]{}%
\providecommand \bibitemNoStop [0]{.\EOS\space}%
\providecommand \EOS [0]{\spacefactor3000\relax}%
\providecommand \BibitemShut  [1]{\csname bibitem#1\endcsname}%
\let\auto@bib@innerbib\@empty
\bibitem [{\citenamefont {Ayral}\ \emph {et~al.}(2023)\citenamefont {Ayral}, \citenamefont {Besserve}, \citenamefont {Lacroix},\ and\ \citenamefont {{Ruiz Guzman}}}]{Ayral2023}%
  \BibitemOpen
  \bibfield  {author} {\bibinfo {author} {\bibfnamefont {T.}~\bibnamefont {Ayral}}, \bibinfo {author} {\bibfnamefont {P.}~\bibnamefont {Besserve}}, \bibinfo {author} {\bibfnamefont {D.}~\bibnamefont {Lacroix}},\ and\ \bibinfo {author} {\bibfnamefont {E.~A.}\ \bibnamefont {{Ruiz Guzman}}},\ }\bibfield  {title} {\bibinfo {title} {{Quantum computing with and for many-body physics}},\ }\href {https://doi.org/10.1140/epja/s10050-023-01141-1} {\bibfield  {journal} {\bibinfo  {journal} {The European Physical Journal A}\ }\textbf {\bibinfo {volume} {59}},\ \bibinfo {pages} {227} (\bibinfo {year} {2023})},\ \Eprint {https://arxiv.org/abs/2303.04850} {arXiv:2303.04850} \BibitemShut {NoStop}%
\bibitem [{\citenamefont {Tokura}\ \emph {et~al.}(2017)\citenamefont {Tokura}, \citenamefont {Kawasaki},\ and\ \citenamefont {Nagaosa}}]{tokura_emergent_2017}%
  \BibitemOpen
  \bibfield  {author} {\bibinfo {author} {\bibfnamefont {Y.}~\bibnamefont {Tokura}}, \bibinfo {author} {\bibfnamefont {M.}~\bibnamefont {Kawasaki}},\ and\ \bibinfo {author} {\bibfnamefont {N.}~\bibnamefont {Nagaosa}},\ }\bibfield  {title} {\bibinfo {title} {Emergent functions of quantum materials},\ }\href {https://doi.org/10.1038/nphys4274} {\bibfield  {journal} {\bibinfo  {journal} {Nature Physics}\ }\textbf {\bibinfo {volume} {13}},\ \bibinfo {pages} {1056} (\bibinfo {year} {2017})},\ \bibinfo {note} {publisher: Nature Publishing Group}\BibitemShut {NoStop}%
\bibitem [{\citenamefont {Cai}\ \emph {et~al.}(2023)\citenamefont {Cai}, \citenamefont {Babbush}, \citenamefont {Benjamin}, \citenamefont {Endo}, \citenamefont {Huggins}, \citenamefont {Li}, \citenamefont {McClean},\ and\ \citenamefont {O'Brien}}]{Cai2022}%
  \BibitemOpen
  \bibfield  {author} {\bibinfo {author} {\bibfnamefont {Z.}~\bibnamefont {Cai}}, \bibinfo {author} {\bibfnamefont {R.}~\bibnamefont {Babbush}}, \bibinfo {author} {\bibfnamefont {S.~C.}\ \bibnamefont {Benjamin}}, \bibinfo {author} {\bibfnamefont {S.}~\bibnamefont {Endo}}, \bibinfo {author} {\bibfnamefont {W.~J.}\ \bibnamefont {Huggins}}, \bibinfo {author} {\bibfnamefont {Y.}~\bibnamefont {Li}}, \bibinfo {author} {\bibfnamefont {J.~R.}\ \bibnamefont {McClean}},\ and\ \bibinfo {author} {\bibfnamefont {T.~E.}\ \bibnamefont {O'Brien}},\ }\bibfield  {title} {\bibinfo {title} {{Quantum error mitigation}},\ }\href {https://doi.org/10.1103/RevModPhys.95.045005} {\bibfield  {journal} {\bibinfo  {journal} {Reviews of Modern Physics}\ }\textbf {\bibinfo {volume} {95}},\ \bibinfo {pages} {045005} (\bibinfo {year} {2023})},\ \Eprint {https://arxiv.org/abs/2210.00921} {arXiv:2210.00921} \BibitemShut {NoStop}%
\bibitem [{\citenamefont {Terhal}(2015)}]{Terhal2015}%
  \BibitemOpen
  \bibfield  {author} {\bibinfo {author} {\bibfnamefont {B.~M.}\ \bibnamefont {Terhal}},\ }\bibfield  {title} {\bibinfo {title} {{Quantum error correction for quantum memories}},\ }\href {https://doi.org/10.1103/RevModPhys.87.307} {\bibfield  {journal} {\bibinfo  {journal} {Reviews of Modern Physics}\ }\textbf {\bibinfo {volume} {87}},\ \bibinfo {pages} {307} (\bibinfo {year} {2015})},\ \Eprint {https://arxiv.org/abs/1302.3428} {arXiv:1302.3428} \BibitemShut {NoStop}%
\bibitem [{\citenamefont {Pocklington}\ \emph {et~al.}(2022)\citenamefont {Pocklington}, \citenamefont {Wang}, \citenamefont {Yanay},\ and\ \citenamefont {Clerk}}]{pocklington_stabilizing_2022}%
  \BibitemOpen
  \bibfield  {author} {\bibinfo {author} {\bibfnamefont {A.}~\bibnamefont {Pocklington}}, \bibinfo {author} {\bibfnamefont {Y.-X.}\ \bibnamefont {Wang}}, \bibinfo {author} {\bibfnamefont {Y.}~\bibnamefont {Yanay}},\ and\ \bibinfo {author} {\bibfnamefont {A.~A.}\ \bibnamefont {Clerk}},\ }\bibfield  {title} {\bibinfo {title} {Stabilizing volume-law entangled states of fermions and qubits using local dissipation},\ }\href {https://doi.org/10.1103/PhysRevB.105.L140301} {\bibfield  {journal} {\bibinfo  {journal} {Physical Review B}\ }\textbf {\bibinfo {volume} {105}},\ \bibinfo {pages} {L140301} (\bibinfo {year} {2022})},\ \bibinfo {note} {publisher: American Physical Society}\BibitemShut {NoStop}%
\bibitem [{\citenamefont {Harrington}\ \emph {et~al.}(2022)\citenamefont {Harrington}, \citenamefont {Mueller},\ and\ \citenamefont {Murch}}]{harrington_engineered_2022}%
  \BibitemOpen
  \bibfield  {author} {\bibinfo {author} {\bibfnamefont {P.~M.}\ \bibnamefont {Harrington}}, \bibinfo {author} {\bibfnamefont {E.~J.}\ \bibnamefont {Mueller}},\ and\ \bibinfo {author} {\bibfnamefont {K.~W.}\ \bibnamefont {Murch}},\ }\bibfield  {title} {\bibinfo {title} {Engineered dissipation for quantum information science},\ }\href {https://doi.org/10.1038/s42254-022-00494-8} {\bibfield  {journal} {\bibinfo  {journal} {Nature Reviews Physics}\ }\textbf {\bibinfo {volume} {4}},\ \bibinfo {pages} {660} (\bibinfo {year} {2022})},\ \bibinfo {note} {number: 10 Publisher: Nature Publishing Group}\BibitemShut {NoStop}%
\bibitem [{\citenamefont {Chen}\ \emph {et~al.}(2023)\citenamefont {Chen}, \citenamefont {Huang}, \citenamefont {Preskill},\ and\ \citenamefont {Zhou}}]{chen_local_2023}%
  \BibitemOpen
  \bibfield  {author} {\bibinfo {author} {\bibfnamefont {C.-F.}\ \bibnamefont {Chen}}, \bibinfo {author} {\bibfnamefont {H.-Y.}\ \bibnamefont {Huang}}, \bibinfo {author} {\bibfnamefont {J.}~\bibnamefont {Preskill}},\ and\ \bibinfo {author} {\bibfnamefont {L.}~\bibnamefont {Zhou}},\ }\href {https://doi.org/10.48550/arXiv.2309.16596} {\bibinfo {title} {Local minima in quantum systems}} (\bibinfo {year} {2023}),\ \bibinfo {note} {arXiv:2309.16596 [cond-mat, physics:math-ph, physics:quant-ph]}\BibitemShut {NoStop}%
\bibitem [{\citenamefont {Lepp\"akangas}\ \emph {et~al.}(2023)\citenamefont {Lepp\"akangas}, \citenamefont {Vogt}, \citenamefont {Fratus}, \citenamefont {Bark}, \citenamefont {Vaitkus}, \citenamefont {Stadler}, \citenamefont {Reiner}, \citenamefont {Zanker},\ and\ \citenamefont {Marthaler}}]{leppakangas_quantum_2023}%
  \BibitemOpen
  \bibfield  {author} {\bibinfo {author} {\bibfnamefont {J.}~\bibnamefont {Lepp\"akangas}}, \bibinfo {author} {\bibfnamefont {N.}~\bibnamefont {Vogt}}, \bibinfo {author} {\bibfnamefont {K.~R.}\ \bibnamefont {Fratus}}, \bibinfo {author} {\bibfnamefont {K.}~\bibnamefont {Bark}}, \bibinfo {author} {\bibfnamefont {J.~A.}\ \bibnamefont {Vaitkus}}, \bibinfo {author} {\bibfnamefont {P.}~\bibnamefont {Stadler}}, \bibinfo {author} {\bibfnamefont {J.-M.}\ \bibnamefont {Reiner}}, \bibinfo {author} {\bibfnamefont {S.}~\bibnamefont {Zanker}},\ and\ \bibinfo {author} {\bibfnamefont {M.}~\bibnamefont {Marthaler}},\ }\href {https://doi.org/10.48550/arXiv.2210.12138} {\bibinfo {title} {A quantum algorithm for solving open system dynamics on quantum computers using noise}} (\bibinfo {year} {2023}),\ \bibinfo {note} {arXiv:2210.12138 [quant-ph]}\BibitemShut {NoStop}%
\bibitem [{\citenamefont {Rost}\ \emph {et~al.}(2020)\citenamefont {Rost}, \citenamefont {Jones}, \citenamefont {Vyushkova}, \citenamefont {Ali}, \citenamefont {Cullip}, \citenamefont {Vyushkov},\ and\ \citenamefont {Nabrzyski}}]{rost_simulation_2020}%
  \BibitemOpen
  \bibfield  {author} {\bibinfo {author} {\bibfnamefont {B.}~\bibnamefont {Rost}}, \bibinfo {author} {\bibfnamefont {B.}~\bibnamefont {Jones}}, \bibinfo {author} {\bibfnamefont {M.}~\bibnamefont {Vyushkova}}, \bibinfo {author} {\bibfnamefont {A.}~\bibnamefont {Ali}}, \bibinfo {author} {\bibfnamefont {C.}~\bibnamefont {Cullip}}, \bibinfo {author} {\bibfnamefont {A.}~\bibnamefont {Vyushkov}},\ and\ \bibinfo {author} {\bibfnamefont {J.}~\bibnamefont {Nabrzyski}},\ }\href {https://doi.org/10.48550/arXiv.2001.00794} {\bibinfo {title} {Simulation of {Thermal} {Relaxation} in {Spin} {Chemistry} {Systems} on a {Quantum} {Computer} {Using} {Inherent} {Qubit} {Decoherence}}} (\bibinfo {year} {2020}),\ \bibinfo {note} {arXiv:2001.00794 [physics, physics:quant-ph]}\BibitemShut {NoStop}%
\bibitem [{\citenamefont {Sun}\ \emph {et~al.}(2021)\citenamefont {Sun}, \citenamefont {Shih},\ and\ \citenamefont {Cheng}}]{sun_efficient_2021}%
  \BibitemOpen
  \bibfield  {author} {\bibinfo {author} {\bibfnamefont {S.}~\bibnamefont {Sun}}, \bibinfo {author} {\bibfnamefont {L.-C.}\ \bibnamefont {Shih}},\ and\ \bibinfo {author} {\bibfnamefont {Y.-C.}\ \bibnamefont {Cheng}},\ }\href {https://doi.org/10.48550/arXiv.2106.12882} {\bibinfo {title} {Efficient {Quantum} {Simulation} of {Open} {Quantum} {System} {Dynamics} on {Noisy} {Quantum} {Computers}}} (\bibinfo {year} {2021}),\ \bibinfo {note} {arXiv:2106.12882 [quant-ph]}\BibitemShut {NoStop}%
\bibitem [{\citenamefont {Georges}\ \emph {et~al.}(1996)\citenamefont {Georges}, \citenamefont {Kotliar}, \citenamefont {Krauth},\ and\ \citenamefont {Rozenberg}}]{Georges1996}%
  \BibitemOpen
  \bibfield  {author} {\bibinfo {author} {\bibfnamefont {A.}~\bibnamefont {Georges}}, \bibinfo {author} {\bibfnamefont {G.}~\bibnamefont {Kotliar}}, \bibinfo {author} {\bibfnamefont {W.}~\bibnamefont {Krauth}},\ and\ \bibinfo {author} {\bibfnamefont {M.~J.}\ \bibnamefont {Rozenberg}},\ }\bibfield  {title} {\bibinfo {title} {{Dynamical mean-field theory of strongly correlated fermion systems and the limit of infinite dimensions}},\ }\href {https://doi.org/10.1103/RevModPhys.68.13} {\bibfield  {journal} {\bibinfo  {journal} {Reviews of Modern Physics}\ }\textbf {\bibinfo {volume} {68}},\ \bibinfo {pages} {13} (\bibinfo {year} {1996})}\BibitemShut {NoStop}%
\bibitem [{\citenamefont {Gull}\ \emph {et~al.}(2011)\citenamefont {Gull}, \citenamefont {Millis}, \citenamefont {Lichtenstein}, \citenamefont {Rubtsov}, \citenamefont {Troyer},\ and\ \citenamefont {Werner}}]{gull_continuous-time_2011}%
  \BibitemOpen
  \bibfield  {author} {\bibinfo {author} {\bibfnamefont {E.}~\bibnamefont {Gull}}, \bibinfo {author} {\bibfnamefont {A.~J.}\ \bibnamefont {Millis}}, \bibinfo {author} {\bibfnamefont {A.~I.}\ \bibnamefont {Lichtenstein}}, \bibinfo {author} {\bibfnamefont {A.~N.}\ \bibnamefont {Rubtsov}}, \bibinfo {author} {\bibfnamefont {M.}~\bibnamefont {Troyer}},\ and\ \bibinfo {author} {\bibfnamefont {P.}~\bibnamefont {Werner}},\ }\bibfield  {title} {\bibinfo {title} {Continuous-time {Monte} {Carlo} methods for quantum impurity models},\ }\href {https://doi.org/10.1103/RevModPhys.83.349} {\bibfield  {journal} {\bibinfo  {journal} {Reviews of Modern Physics}\ }\textbf {\bibinfo {volume} {83}},\ \bibinfo {pages} {349} (\bibinfo {year} {2011})},\ \bibinfo {note} {publisher: American Physical Society}\BibitemShut {NoStop}%
\bibitem [{\citenamefont {{Simons Collaboration on the Many-Electron Problem}}\ \emph {et~al.}(2015)\citenamefont {{Simons Collaboration on the Many-Electron Problem}}, \citenamefont {LeBlanc}, \citenamefont {Antipov}, \citenamefont {Becca}, \citenamefont {Bulik}, \citenamefont {Chan}, \citenamefont {Chung}, \citenamefont {Deng}, \citenamefont {Ferrero}, \citenamefont {Henderson}, \citenamefont {Jim\'enez-Hoyos}, \citenamefont {Kozik}, \citenamefont {Liu}, \citenamefont {Millis}, \citenamefont {Prokof'ev}, \citenamefont {Qin}, \citenamefont {Scuseria}, \citenamefont {Shi}, \citenamefont {Svistunov}, \citenamefont {Tocchio}, \citenamefont {Tupitsyn}, \citenamefont {White}, \citenamefont {Zhang}, \citenamefont {Zheng}, \citenamefont {Zhu},\ and\ \citenamefont {Gull}}]{simons_collaboration_2015}%
  \BibitemOpen
  \bibfield  {author} {\bibinfo {author} {\bibnamefont {{Simons Collaboration on the Many-Electron Problem}}}, \bibinfo {author} {\bibfnamefont {J.~P.~F.}\ \bibnamefont {LeBlanc}}, \bibinfo {author} {\bibfnamefont {A.~E.}\ \bibnamefont {Antipov}}, \bibinfo {author} {\bibfnamefont {F.}~\bibnamefont {Becca}}, \bibinfo {author} {\bibfnamefont {I.~W.}\ \bibnamefont {Bulik}}, \bibinfo {author} {\bibfnamefont {G.~K.-L.}\ \bibnamefont {Chan}}, \bibinfo {author} {\bibfnamefont {C.-M.}\ \bibnamefont {Chung}}, \bibinfo {author} {\bibfnamefont {Y.}~\bibnamefont {Deng}}, \bibinfo {author} {\bibfnamefont {M.}~\bibnamefont {Ferrero}}, \bibinfo {author} {\bibfnamefont {T.~M.}\ \bibnamefont {Henderson}}, \bibinfo {author} {\bibfnamefont {C.~A.}\ \bibnamefont {Jim\'enez-Hoyos}}, \bibinfo {author} {\bibfnamefont {E.}~\bibnamefont {Kozik}}, \bibinfo {author} {\bibfnamefont {X.-W.}\ \bibnamefont {Liu}}, \bibinfo {author} {\bibfnamefont {A.~J.}\ \bibnamefont {Millis}}, \bibinfo {author} {\bibfnamefont {N.~V.}\ \bibnamefont
  {Prokof'ev}}, \bibinfo {author} {\bibfnamefont {M.}~\bibnamefont {Qin}}, \bibinfo {author} {\bibfnamefont {G.~E.}\ \bibnamefont {Scuseria}}, \bibinfo {author} {\bibfnamefont {H.}~\bibnamefont {Shi}}, \bibinfo {author} {\bibfnamefont {B.~V.}\ \bibnamefont {Svistunov}}, \bibinfo {author} {\bibfnamefont {L.~F.}\ \bibnamefont {Tocchio}}, \bibinfo {author} {\bibfnamefont {I.~S.}\ \bibnamefont {Tupitsyn}}, \bibinfo {author} {\bibfnamefont {S.~R.}\ \bibnamefont {White}}, \bibinfo {author} {\bibfnamefont {S.}~\bibnamefont {Zhang}}, \bibinfo {author} {\bibfnamefont {B.-X.}\ \bibnamefont {Zheng}}, \bibinfo {author} {\bibfnamefont {Z.}~\bibnamefont {Zhu}},\ and\ \bibinfo {author} {\bibfnamefont {E.}~\bibnamefont {Gull}},\ }\bibfield  {title} {\bibinfo {title} {Solutions of the {Two}-{Dimensional} {Hubbard} {Model}: {Benchmarks} and {Results} from a {Wide} {Range} of {Numerical} {Algorithms}},\ }\href {https://doi.org/10.1103/PhysRevX.5.041041} {\bibfield  {journal} {\bibinfo  {journal} {Physical Review X}\ }\textbf
  {\bibinfo {volume} {5}},\ \bibinfo {pages} {041041} (\bibinfo {year} {2015})},\ \bibinfo {note} {publisher: American Physical Society}\BibitemShut {NoStop}%
\bibitem [{\citenamefont {Wolf}\ \emph {et~al.}(2014)\citenamefont {Wolf}, \citenamefont {McCulloch},\ and\ \citenamefont {Schollw\"ock}}]{wolf_solving_2014}%
  \BibitemOpen
  \bibfield  {author} {\bibinfo {author} {\bibfnamefont {F.~A.}\ \bibnamefont {Wolf}}, \bibinfo {author} {\bibfnamefont {I.~P.}\ \bibnamefont {McCulloch}},\ and\ \bibinfo {author} {\bibfnamefont {U.}~\bibnamefont {Schollw\"ock}},\ }\bibfield  {title} {\bibinfo {title} {Solving nonequilibrium dynamical mean-field theory using matrix product states},\ }\href {https://doi.org/10.1103/PhysRevB.90.235131} {\bibfield  {journal} {\bibinfo  {journal} {Physical Review B}\ }\textbf {\bibinfo {volume} {90}},\ \bibinfo {pages} {235131} (\bibinfo {year} {2014})},\ \bibinfo {note} {publisher: American Physical Society}\BibitemShut {NoStop}%
\bibitem [{\citenamefont {Bauernfeind}\ \emph {et~al.}(2017)\citenamefont {Bauernfeind}, \citenamefont {Zingl}, \citenamefont {Triebl}, \citenamefont {Aichhorn},\ and\ \citenamefont {Evertz}}]{bauernfeind_fork_2017}%
  \BibitemOpen
  \bibfield  {author} {\bibinfo {author} {\bibfnamefont {D.}~\bibnamefont {Bauernfeind}}, \bibinfo {author} {\bibfnamefont {M.}~\bibnamefont {Zingl}}, \bibinfo {author} {\bibfnamefont {R.}~\bibnamefont {Triebl}}, \bibinfo {author} {\bibfnamefont {M.}~\bibnamefont {Aichhorn}},\ and\ \bibinfo {author} {\bibfnamefont {H.~G.}\ \bibnamefont {Evertz}},\ }\bibfield  {title} {\bibinfo {title} {Fork {Tensor}-{Product} {States}: {Efficient} {Multiorbital} {Real}-{Time} {DMFT} {Solver}},\ }\href {https://doi.org/10.1103/PhysRevX.7.031013} {\bibfield  {journal} {\bibinfo  {journal} {Physical Review X}\ }\textbf {\bibinfo {volume} {7}},\ \bibinfo {pages} {031013} (\bibinfo {year} {2017})},\ \bibinfo {note} {publisher: American Physical Society}\BibitemShut {NoStop}%
\bibitem [{\citenamefont {Kreula}\ \emph {et~al.}(2016{\natexlab{a}})\citenamefont {Kreula}, \citenamefont {Clark},\ and\ \citenamefont {Jaksch}}]{Kreula2016}%
  \BibitemOpen
  \bibfield  {author} {\bibinfo {author} {\bibfnamefont {J.~M.}\ \bibnamefont {Kreula}}, \bibinfo {author} {\bibfnamefont {S.~R.}\ \bibnamefont {Clark}},\ and\ \bibinfo {author} {\bibfnamefont {D.}~\bibnamefont {Jaksch}},\ }\bibfield  {title} {\bibinfo {title} {{Non-linear quantum-classical scheme to simulate non-equilibrium strongly correlated fermionic many-body dynamics}},\ }\href {https://doi.org/10.1038/srep32940} {\bibfield  {journal} {\bibinfo  {journal} {Scientific Reports}\ }\textbf {\bibinfo {volume} {6}},\ \bibinfo {pages} {32940} (\bibinfo {year} {2016}{\natexlab{a}})},\ \Eprint {https://arxiv.org/abs/1510.05703} {arXiv:1510.05703} \BibitemShut {NoStop}%
\bibitem [{\citenamefont {Kreula}\ \emph {et~al.}(2016{\natexlab{b}})\citenamefont {Kreula}, \citenamefont {Garc{\'{i}}a-{\'{A}}lvarez}, \citenamefont {Lamata}, \citenamefont {Clark}, \citenamefont {Solano},\ and\ \citenamefont {Jaksch}}]{Kreula2016a}%
  \BibitemOpen
  \bibfield  {author} {\bibinfo {author} {\bibfnamefont {J.~M.}\ \bibnamefont {Kreula}}, \bibinfo {author} {\bibfnamefont {L.}~\bibnamefont {Garc{\'{i}}a-{\'{A}}lvarez}}, \bibinfo {author} {\bibfnamefont {L.}~\bibnamefont {Lamata}}, \bibinfo {author} {\bibfnamefont {S.~R.}\ \bibnamefont {Clark}}, \bibinfo {author} {\bibfnamefont {E.}~\bibnamefont {Solano}},\ and\ \bibinfo {author} {\bibfnamefont {D.}~\bibnamefont {Jaksch}},\ }\bibfield  {title} {\bibinfo {title} {{Few-qubit quantum-classical simulation of strongly correlated lattice fermions}},\ }\href {https://doi.org/10.1140/epjqt/s40507-016-0049-1} {\bibfield  {journal} {\bibinfo  {journal} {EPJ Quantum Technology}\ }\textbf {\bibinfo {volume} {3}},\ \bibinfo {pages} {11} (\bibinfo {year} {2016}{\natexlab{b}})},\ \Eprint {https://arxiv.org/abs/1606.04839} {arXiv:1606.04839} \BibitemShut {NoStop}%
\bibitem [{\citenamefont {Bauer}\ \emph {et~al.}(2016)\citenamefont {Bauer}, \citenamefont {Wecker}, \citenamefont {Millis}, \citenamefont {Hastings},\ and\ \citenamefont {Troyer}}]{Bauer2016}%
  \BibitemOpen
  \bibfield  {author} {\bibinfo {author} {\bibfnamefont {B.}~\bibnamefont {Bauer}}, \bibinfo {author} {\bibfnamefont {D.}~\bibnamefont {Wecker}}, \bibinfo {author} {\bibfnamefont {A.~J.}\ \bibnamefont {Millis}}, \bibinfo {author} {\bibfnamefont {M.~B.}\ \bibnamefont {Hastings}},\ and\ \bibinfo {author} {\bibfnamefont {M.}~\bibnamefont {Troyer}},\ }\bibfield  {title} {\bibinfo {title} {{Hybrid Quantum-Classical Approach to Correlated Materials}},\ }\href {https://doi.org/10.1103/PhysRevX.6.031045} {\bibfield  {journal} {\bibinfo  {journal} {Physical Review X}\ }\textbf {\bibinfo {volume} {6}},\ \bibinfo {pages} {031045} (\bibinfo {year} {2016})}\BibitemShut {NoStop}%
\bibitem [{\citenamefont {Rungger}\ \emph {et~al.}(2019)\citenamefont {Rungger}, \citenamefont {Fitzpatrick}, \citenamefont {Chen}, \citenamefont {Alderete}, \citenamefont {Apel}, \citenamefont {Cowtan}, \citenamefont {Patterson}, \citenamefont {Ramo}, \citenamefont {Zhu}, \citenamefont {Nguyen}, \citenamefont {Grant}, \citenamefont {Chretien}, \citenamefont {Wossnig}, \citenamefont {Linke},\ and\ \citenamefont {Duncan}}]{Rungger2019}%
  \BibitemOpen
  \bibfield  {author} {\bibinfo {author} {\bibfnamefont {I.}~\bibnamefont {Rungger}}, \bibinfo {author} {\bibfnamefont {N.}~\bibnamefont {Fitzpatrick}}, \bibinfo {author} {\bibfnamefont {H.}~\bibnamefont {Chen}}, \bibinfo {author} {\bibfnamefont {C.~H.}\ \bibnamefont {Alderete}}, \bibinfo {author} {\bibfnamefont {H.}~\bibnamefont {Apel}}, \bibinfo {author} {\bibfnamefont {A.}~\bibnamefont {Cowtan}}, \bibinfo {author} {\bibfnamefont {A.}~\bibnamefont {Patterson}}, \bibinfo {author} {\bibfnamefont {D.~M.}\ \bibnamefont {Ramo}}, \bibinfo {author} {\bibfnamefont {Y.}~\bibnamefont {Zhu}}, \bibinfo {author} {\bibfnamefont {N.~H.}\ \bibnamefont {Nguyen}}, \bibinfo {author} {\bibfnamefont {E.}~\bibnamefont {Grant}}, \bibinfo {author} {\bibfnamefont {S.}~\bibnamefont {Chretien}}, \bibinfo {author} {\bibfnamefont {L.}~\bibnamefont {Wossnig}}, \bibinfo {author} {\bibfnamefont {N.~M.}\ \bibnamefont {Linke}},\ and\ \bibinfo {author} {\bibfnamefont {R.}~\bibnamefont {Duncan}},\ }\bibfield  {title} {\bibinfo {title}
  {{Dynamical mean field theory algorithm and experiment on quantum computers}},\ }\href {http://arxiv.org/abs/1910.04735} {\ ,\ \bibinfo {pages} {1} (\bibinfo {year} {2019})},\ \Eprint {https://arxiv.org/abs/1910.04735} {arXiv:1910.04735} \BibitemShut {NoStop}%
\bibitem [{\citenamefont {Yao}\ \emph {et~al.}(2021)\citenamefont {Yao}, \citenamefont {Zhang}, \citenamefont {Wang}, \citenamefont {Ho},\ and\ \citenamefont {Orth}}]{Yao2020}%
  \BibitemOpen
  \bibfield  {author} {\bibinfo {author} {\bibfnamefont {Y.}~\bibnamefont {Yao}}, \bibinfo {author} {\bibfnamefont {F.}~\bibnamefont {Zhang}}, \bibinfo {author} {\bibfnamefont {C.-Z.}\ \bibnamefont {Wang}}, \bibinfo {author} {\bibfnamefont {K.-M.}\ \bibnamefont {Ho}},\ and\ \bibinfo {author} {\bibfnamefont {P.~P.}\ \bibnamefont {Orth}},\ }\bibfield  {title} {\bibinfo {title} {{Gutzwiller hybrid quantum-classical computing approach for correlated materials}},\ }\href {https://doi.org/10.1103/PhysRevResearch.3.013184} {\bibfield  {journal} {\bibinfo  {journal} {Physical Review Research}\ }\textbf {\bibinfo {volume} {3}},\ \bibinfo {pages} {013184} (\bibinfo {year} {2021})},\ \Eprint {https://arxiv.org/abs/2003.04211} {arXiv:2003.04211} \BibitemShut {NoStop}%
\bibitem [{\citenamefont {Jaderberg}\ \emph {et~al.}(2020)\citenamefont {Jaderberg}, \citenamefont {Agarwal}, \citenamefont {Leonhardt}, \citenamefont {Kiffner},\ and\ \citenamefont {Jaksch}}]{Jaderberg2020}%
  \BibitemOpen
  \bibfield  {author} {\bibinfo {author} {\bibfnamefont {B.}~\bibnamefont {Jaderberg}}, \bibinfo {author} {\bibfnamefont {A.}~\bibnamefont {Agarwal}}, \bibinfo {author} {\bibfnamefont {K.}~\bibnamefont {Leonhardt}}, \bibinfo {author} {\bibfnamefont {M.}~\bibnamefont {Kiffner}},\ and\ \bibinfo {author} {\bibfnamefont {D.}~\bibnamefont {Jaksch}},\ }\bibfield  {title} {\bibinfo {title} {{Minimum hardware requirements for hybrid quantum–classical DMFT}},\ }\href {https://doi.org/10.1088/2058-9565/ab972b} {\bibfield  {journal} {\bibinfo  {journal} {Quantum Science and Technology}\ }\textbf {\bibinfo {volume} {5}},\ \bibinfo {pages} {034015} (\bibinfo {year} {2020})},\ \Eprint {https://arxiv.org/abs/2002.04612} {arXiv:2002.04612} \BibitemShut {NoStop}%
\bibitem [{\citenamefont {Besserve}\ and\ \citenamefont {Ayral}(2022)}]{Besserve2021}%
  \BibitemOpen
  \bibfield  {author} {\bibinfo {author} {\bibfnamefont {P.}~\bibnamefont {Besserve}}\ and\ \bibinfo {author} {\bibfnamefont {T.}~\bibnamefont {Ayral}},\ }\bibfield  {title} {\bibinfo {title} {{Unraveling correlated material properties with noisy quantum computers: Natural orbitalized variational quantum eigensolving of extended impurity models within a slave-boson approach}},\ }\href {https://doi.org/10.1103/PhysRevB.105.115108} {\bibfield  {journal} {\bibinfo  {journal} {Physical Review B}\ }\textbf {\bibinfo {volume} {105}},\ \bibinfo {pages} {115108} (\bibinfo {year} {2022})},\ \Eprint {https://arxiv.org/abs/2108.10780} {arXiv:2108.10780} \BibitemShut {NoStop}%
\bibitem [{\citenamefont {Garraway}(1997)}]{garraway_nonperturbative_1997}%
  \BibitemOpen
  \bibfield  {author} {\bibinfo {author} {\bibfnamefont {B.~M.}\ \bibnamefont {Garraway}},\ }\bibfield  {title} {\bibinfo {title} {Nonperturbative decay of an atomic system in a cavity},\ }\href {https://doi.org/10.1103/PhysRevA.55.2290} {\bibfield  {journal} {\bibinfo  {journal} {Physical Review A}\ }\textbf {\bibinfo {volume} {55}},\ \bibinfo {pages} {2290} (\bibinfo {year} {1997})},\ \bibinfo {note} {publisher: American Physical Society}\BibitemShut {NoStop}%
\bibitem [{\citenamefont {Kakuyanagi}\ \emph {et~al.}(2007)\citenamefont {Kakuyanagi}, \citenamefont {Meno}, \citenamefont {Saito}, \citenamefont {Nakano}, \citenamefont {Semba}, \citenamefont {Takayanagi}, \citenamefont {Deppe},\ and\ \citenamefont {Shnirman}}]{kakuyanagi_dephasing_2007}%
  \BibitemOpen
  \bibfield  {author} {\bibinfo {author} {\bibfnamefont {K.}~\bibnamefont {Kakuyanagi}}, \bibinfo {author} {\bibfnamefont {T.}~\bibnamefont {Meno}}, \bibinfo {author} {\bibfnamefont {S.}~\bibnamefont {Saito}}, \bibinfo {author} {\bibfnamefont {H.}~\bibnamefont {Nakano}}, \bibinfo {author} {\bibfnamefont {K.}~\bibnamefont {Semba}}, \bibinfo {author} {\bibfnamefont {H.}~\bibnamefont {Takayanagi}}, \bibinfo {author} {\bibfnamefont {F.}~\bibnamefont {Deppe}},\ and\ \bibinfo {author} {\bibfnamefont {A.}~\bibnamefont {Shnirman}},\ }\bibfield  {title} {\bibinfo {title} {Dephasing of a {Superconducting} {Flux} {Qubit}},\ }\href {https://doi.org/10.1103/PhysRevLett.98.047004} {\bibfield  {journal} {\bibinfo  {journal} {Physical Review Letters}\ }\textbf {\bibinfo {volume} {98}},\ \bibinfo {pages} {047004} (\bibinfo {year} {2007})},\ \bibinfo {note} {publisher: American Physical Society}\BibitemShut {NoStop}%
\bibitem [{\citenamefont {Kubica}\ \emph {et~al.}(2023)\citenamefont {Kubica}, \citenamefont {Haim}, \citenamefont {Vaknin}, \citenamefont {Levine}, \citenamefont {Brand\~ao},\ and\ \citenamefont {Retzker}}]{kubica_erasure_2023}%
  \BibitemOpen
  \bibfield  {author} {\bibinfo {author} {\bibfnamefont {A.}~\bibnamefont {Kubica}}, \bibinfo {author} {\bibfnamefont {A.}~\bibnamefont {Haim}}, \bibinfo {author} {\bibfnamefont {Y.}~\bibnamefont {Vaknin}}, \bibinfo {author} {\bibfnamefont {H.}~\bibnamefont {Levine}}, \bibinfo {author} {\bibfnamefont {F.}~\bibnamefont {Brand\~ao}},\ and\ \bibinfo {author} {\bibfnamefont {A.}~\bibnamefont {Retzker}},\ }\bibfield  {title} {\bibinfo {title} {Erasure {Qubits}: {Overcoming} the {$T_1$} {Limit} in {Superconducting} {Circuits}},\ }\href {https://doi.org/10.1103/PhysRevX.13.041022} {\bibfield  {journal} {\bibinfo  {journal} {Physical Review X}\ }\textbf {\bibinfo {volume} {13}},\ \bibinfo {pages} {041022} (\bibinfo {year} {2023})},\ \bibinfo {note} {publisher: American Physical Society}\BibitemShut {NoStop}%
\bibitem [{\citenamefont {Temme}\ \emph {et~al.}(2017)\citenamefont {Temme}, \citenamefont {Bravyi},\ and\ \citenamefont {Gambetta}}]{temme_error_2017}%
  \BibitemOpen
  \bibfield  {author} {\bibinfo {author} {\bibfnamefont {K.}~\bibnamefont {Temme}}, \bibinfo {author} {\bibfnamefont {S.}~\bibnamefont {Bravyi}},\ and\ \bibinfo {author} {\bibfnamefont {J.~M.}\ \bibnamefont {Gambetta}},\ }\bibfield  {title} {\bibinfo {title} {Error {Mitigation} for {Short}-{Depth} {Quantum} {Circuits}},\ }\href {https://doi.org/10.1103/PhysRevLett.119.180509} {\bibfield  {journal} {\bibinfo  {journal} {Physical Review Letters}\ }\textbf {\bibinfo {volume} {119}},\ \bibinfo {pages} {180509} (\bibinfo {year} {2017})},\ \bibinfo {note} {publisher: American Physical Society}\BibitemShut {NoStop}%
\bibitem [{\citenamefont {Bultrini}\ \emph {et~al.}(2023)\citenamefont {Bultrini}, \citenamefont {Wang}, \citenamefont {Czarnik}, \citenamefont {Gordon}, \citenamefont {Cerezo}, \citenamefont {Coles},\ and\ \citenamefont {Cincio}}]{bultrini_battle_2023}%
  \BibitemOpen
  \bibfield  {author} {\bibinfo {author} {\bibfnamefont {D.}~\bibnamefont {Bultrini}}, \bibinfo {author} {\bibfnamefont {S.}~\bibnamefont {Wang}}, \bibinfo {author} {\bibfnamefont {P.}~\bibnamefont {Czarnik}}, \bibinfo {author} {\bibfnamefont {M.~H.}\ \bibnamefont {Gordon}}, \bibinfo {author} {\bibfnamefont {M.}~\bibnamefont {Cerezo}}, \bibinfo {author} {\bibfnamefont {P.~J.}\ \bibnamefont {Coles}},\ and\ \bibinfo {author} {\bibfnamefont {L.}~\bibnamefont {Cincio}},\ }\bibfield  {title} {\bibinfo {title} {The battle of clean and dirty qubits in the era of partial error correction},\ }\href {https://doi.org/10.22331/q-2023-07-13-1060} {\bibfield  {journal} {\bibinfo  {journal} {Quantum}\ }\textbf {\bibinfo {volume} {7}},\ \bibinfo {pages} {1060} (\bibinfo {year} {2023})},\ \bibinfo {note} {arXiv:2205.13454 [quant-ph]}\BibitemShut {NoStop}%
\bibitem [{\citenamefont {Koukoulekidis}\ \emph {et~al.}(2023)\citenamefont {Koukoulekidis}, \citenamefont {Wang}, \citenamefont {O'Leary}, \citenamefont {Bultrini}, \citenamefont {Cincio},\ and\ \citenamefont {Czarnik}}]{koukoulekidis_framework_2023}%
  \BibitemOpen
  \bibfield  {author} {\bibinfo {author} {\bibfnamefont {N.}~\bibnamefont {Koukoulekidis}}, \bibinfo {author} {\bibfnamefont {S.}~\bibnamefont {Wang}}, \bibinfo {author} {\bibfnamefont {T.}~\bibnamefont {O'Leary}}, \bibinfo {author} {\bibfnamefont {D.}~\bibnamefont {Bultrini}}, \bibinfo {author} {\bibfnamefont {L.}~\bibnamefont {Cincio}},\ and\ \bibinfo {author} {\bibfnamefont {P.}~\bibnamefont {Czarnik}},\ }\href {https://doi.org/10.48550/arXiv.2306.15531} {\bibinfo {title} {A framework of partial error correction for intermediate-scale quantum computers}} (\bibinfo {year} {2023}),\ \bibinfo {note} {arXiv:2306.15531 [quant-ph]}\BibitemShut {NoStop}%
\bibitem [{\citenamefont {Kamenev}(2023)}]{kamenev2023}%
  \BibitemOpen
  \bibfield  {author} {\bibinfo {author} {\bibfnamefont {A.}~\bibnamefont {Kamenev}},\ }\href@noop {} {{\selectlanguage {English}\emph {\bibinfo {title} {Field {Theory} of {Non}-{Equilibrium} {Systems}}}}},\ \bibinfo {edition} {2nd}\ ed.\ (\bibinfo  {publisher} {Cambridge University Press},\ \bibinfo {address} {Cambridge},\ \bibinfo {year} {2023})\BibitemShut {NoStop}%
\bibitem [{\citenamefont {Emery}\ and\ \citenamefont {Kivelson}(1992)}]{emery_mapping_1992}%
  \BibitemOpen
  \bibfield  {author} {\bibinfo {author} {\bibfnamefont {V.~J.}\ \bibnamefont {Emery}}\ and\ \bibinfo {author} {\bibfnamefont {S.}~\bibnamefont {Kivelson}},\ }\bibfield  {title} {\bibinfo {title} {Mapping of the two-channel {Kondo} problem to a resonant-level model},\ }\href {https://doi.org/10.1103/PhysRevB.46.10812} {\bibfield  {journal} {\bibinfo  {journal} {Physical Review B}\ }\textbf {\bibinfo {volume} {46}},\ \bibinfo {pages} {10812} (\bibinfo {year} {1992})},\ \bibinfo {note} {publisher: American Physical Society}\BibitemShut {NoStop}%
\bibitem [{\citenamefont {Gramsch}\ \emph {et~al.}(2013)\citenamefont {Gramsch}, \citenamefont {Balzer}, \citenamefont {Eckstein},\ and\ \citenamefont {Kollar}}]{gramsch_hamiltonian-based_2013}%
  \BibitemOpen
  \bibfield  {author} {\bibinfo {author} {\bibfnamefont {C.}~\bibnamefont {Gramsch}}, \bibinfo {author} {\bibfnamefont {K.}~\bibnamefont {Balzer}}, \bibinfo {author} {\bibfnamefont {M.}~\bibnamefont {Eckstein}},\ and\ \bibinfo {author} {\bibfnamefont {M.}~\bibnamefont {Kollar}},\ }\bibfield  {title} {\bibinfo {title} {Hamiltonian-based impurity solver for nonequilibrium dynamical mean-field theory},\ }\href {https://doi.org/10.1103/PhysRevB.88.235106} {\bibfield  {journal} {\bibinfo  {journal} {Physical Review B}\ }\textbf {\bibinfo {volume} {88}},\ \bibinfo {pages} {235106} (\bibinfo {year} {2013})},\ \bibinfo {note} {publisher: American Physical Society}\BibitemShut {NoStop}%
\bibitem [{\citenamefont {Tamascelli}\ \emph {et~al.}(2018)\citenamefont {Tamascelli}, \citenamefont {Smirne}, \citenamefont {Huelga},\ and\ \citenamefont {Plenio}}]{tamascelli_nonperturbative_2018}%
  \BibitemOpen
  \bibfield  {author} {\bibinfo {author} {\bibfnamefont {D.}~\bibnamefont {Tamascelli}}, \bibinfo {author} {\bibfnamefont {A.}~\bibnamefont {Smirne}}, \bibinfo {author} {\bibfnamefont {S.~F.}\ \bibnamefont {Huelga}},\ and\ \bibinfo {author} {\bibfnamefont {M.~B.}\ \bibnamefont {Plenio}},\ }\bibfield  {title} {\bibinfo {title} {Nonperturbative {Treatment} of non-{Markovian} {Dynamics} of {Open} {Quantum} {Systems}},\ }\href {https://doi.org/10.1103/PhysRevLett.120.030402} {\bibfield  {journal} {\bibinfo  {journal} {Physical Review Letters}\ }\textbf {\bibinfo {volume} {120}},\ \bibinfo {pages} {030402} (\bibinfo {year} {2018})},\ \bibinfo {note} {publisher: American Physical Society}\BibitemShut {NoStop}%
\bibitem [{\citenamefont {Xu}\ \emph {et~al.}(2017)\citenamefont {Xu}, \citenamefont {Liu}, \citenamefont {Zhang},\ and\ \citenamefont {Yan}}]{xu_theory_2017}%
  \BibitemOpen
  \bibfield  {author} {\bibinfo {author} {\bibfnamefont {R.-X.}\ \bibnamefont {Xu}}, \bibinfo {author} {\bibfnamefont {Y.}~\bibnamefont {Liu}}, \bibinfo {author} {\bibfnamefont {H.-D.}\ \bibnamefont {Zhang}},\ and\ \bibinfo {author} {\bibfnamefont {Y.}~\bibnamefont {Yan}},\ }\bibfield  {title} {\bibinfo {title} {Theory of quantum dissipation in a class of non-{Gaussian} environments},\ }\href {https://doi.org/10.1063/1674-0068/30/cjcp1706123} {\bibfield  {journal} {\bibinfo  {journal} {Chinese Journal of Chemical Physics}\ }\textbf {\bibinfo {volume} {30}},\ \bibinfo {pages} {395} (\bibinfo {year} {2017})},\ \bibinfo {note} {arXiv:1608.07774 [physics]}\BibitemShut {NoStop}%
\bibitem [{\citenamefont {Lambert}\ \emph {et~al.}(2019)\citenamefont {Lambert}, \citenamefont {Ahmed}, \citenamefont {Cirio},\ and\ \citenamefont {Nori}}]{lambert_modelling_2019}%
  \BibitemOpen
  \bibfield  {author} {\bibinfo {author} {\bibfnamefont {N.}~\bibnamefont {Lambert}}, \bibinfo {author} {\bibfnamefont {S.}~\bibnamefont {Ahmed}}, \bibinfo {author} {\bibfnamefont {M.}~\bibnamefont {Cirio}},\ and\ \bibinfo {author} {\bibfnamefont {F.}~\bibnamefont {Nori}},\ }\bibfield  {title} {\bibinfo {title} {Modelling the ultra-strongly coupled spin-boson model with unphysical modes},\ }\href {https://doi.org/10.1038/s41467-019-11656-1} {\bibfield  {journal} {\bibinfo  {journal} {Nature Communications}\ }\textbf {\bibinfo {volume} {10}},\ \bibinfo {pages} {3721} (\bibinfo {year} {2019})},\ \bibinfo {note} {number: 1 Publisher: Nature Publishing Group}\BibitemShut {NoStop}%
\bibitem [{\citenamefont {Pleasance}\ \emph {et~al.}(2020)\citenamefont {Pleasance}, \citenamefont {Garraway},\ and\ \citenamefont {Petruccione}}]{pleasance_generalized_2020}%
  \BibitemOpen
  \bibfield  {author} {\bibinfo {author} {\bibfnamefont {G.}~\bibnamefont {Pleasance}}, \bibinfo {author} {\bibfnamefont {B.~M.}\ \bibnamefont {Garraway}},\ and\ \bibinfo {author} {\bibfnamefont {F.}~\bibnamefont {Petruccione}},\ }\bibfield  {title} {\bibinfo {title} {Generalized theory of pseudomodes for exact descriptions of non-{Markovian} quantum processes},\ }\href {https://doi.org/10.1103/PhysRevResearch.2.043058} {\bibfield  {journal} {\bibinfo  {journal} {Physical Review Research}\ }\textbf {\bibinfo {volume} {2}},\ \bibinfo {pages} {043058} (\bibinfo {year} {2020})},\ \bibinfo {note} {publisher: American Physical Society}\BibitemShut {NoStop}%
\bibitem [{\citenamefont {Pleasance}\ and\ \citenamefont {Petruccione}(2021)}]{pleasance_pseudomode_2021}%
  \BibitemOpen
  \bibfield  {author} {\bibinfo {author} {\bibfnamefont {G.}~\bibnamefont {Pleasance}}\ and\ \bibinfo {author} {\bibfnamefont {F.}~\bibnamefont {Petruccione}},\ }\href {https://doi.org/10.48550/arXiv.2108.05755} {\bibinfo {title} {Pseudomode description of general open quantum system dynamics: non-perturbative master equation for the spin-boson model}} (\bibinfo {year} {2021}),\ \bibinfo {note} {arXiv:2108.05755 [quant-ph]}\BibitemShut {NoStop}%
\bibitem [{\citenamefont {Chen}\ and\ \citenamefont {Franco}(2024)}]{chen_bexcitonics_2024}%
  \BibitemOpen
  \bibfield  {author} {\bibinfo {author} {\bibfnamefont {X.}~\bibnamefont {Chen}}\ and\ \bibinfo {author} {\bibfnamefont {I.}~\bibnamefont {Franco}},\ }\href {http://arxiv.org/abs/2401.11049} {\bibinfo {title} {Bexcitonics: {Quasi}-particle approach to open quantum dynamics}} (\bibinfo {year} {2024}),\ \bibinfo {note} {arXiv:2401.11049 [physics, physics:quant-ph]}\BibitemShut {NoStop}%
\bibitem [{\citenamefont {Menczel}\ \emph {et~al.}(2024)\citenamefont {Menczel}, \citenamefont {Funo}, \citenamefont {Cirio}, \citenamefont {Lambert},\ and\ \citenamefont {Nori}}]{menczel_non-hermitian_2024}%
  \BibitemOpen
  \bibfield  {author} {\bibinfo {author} {\bibfnamefont {P.}~\bibnamefont {Menczel}}, \bibinfo {author} {\bibfnamefont {K.}~\bibnamefont {Funo}}, \bibinfo {author} {\bibfnamefont {M.}~\bibnamefont {Cirio}}, \bibinfo {author} {\bibfnamefont {N.}~\bibnamefont {Lambert}},\ and\ \bibinfo {author} {\bibfnamefont {F.}~\bibnamefont {Nori}},\ }\href {http://arxiv.org/abs/2401.11830} {\bibinfo {title} {Non-{Hermitian} {Pseudomodes} for {Strongly} {Coupled} {Open} {Quantum} {Systems}: {Unravelings}, {Correlations} and {Thermodynamics}}} (\bibinfo {year} {2024}),\ \bibinfo {note} {arXiv:2401.11830 [cond-mat, physics:quant-ph]}\BibitemShut {NoStop}%
\bibitem [{\citenamefont {Lacroix}\ \emph {et~al.}(2015)\citenamefont {Lacroix}, \citenamefont {Sargsyan}, \citenamefont {Adamian},\ and\ \citenamefont {Antonenko}}]{lacroix_description_2015}%
  \BibitemOpen
  \bibfield  {author} {\bibinfo {author} {\bibfnamefont {D.}~\bibnamefont {Lacroix}}, \bibinfo {author} {\bibfnamefont {V.}~\bibnamefont {Sargsyan}}, \bibinfo {author} {\bibfnamefont {G.}~\bibnamefont {Adamian}},\ and\ \bibinfo {author} {\bibfnamefont {N.}~\bibnamefont {Antonenko}},\ }\bibfield  {title} {\bibinfo {title} {Description of non-{Markovian} effect in open quantum system with the discretized environment method},\ }\href {https://doi.org/10.1140/epjb/e2015-60052-3} {\bibfield  {journal} {\bibinfo  {journal} {The European Physical Journal B}\ }\textbf {\bibinfo {volume} {88}},\ \bibinfo {pages} {89} (\bibinfo {year} {2015})}\BibitemShut {NoStop}%
\bibitem [{\citenamefont {Lacroix}\ \emph {et~al.}(2020)\citenamefont {Lacroix}, \citenamefont {Sargsyan}, \citenamefont {Adamian}, \citenamefont {Antonenko},\ and\ \citenamefont {Hovhannisyan}}]{lacroix_non-markovian_2020}%
  \BibitemOpen
  \bibfield  {author} {\bibinfo {author} {\bibfnamefont {D.}~\bibnamefont {Lacroix}}, \bibinfo {author} {\bibfnamefont {V.~V.}\ \bibnamefont {Sargsyan}}, \bibinfo {author} {\bibfnamefont {G.~G.}\ \bibnamefont {Adamian}}, \bibinfo {author} {\bibfnamefont {N.~V.}\ \bibnamefont {Antonenko}},\ and\ \bibinfo {author} {\bibfnamefont {A.~A.}\ \bibnamefont {Hovhannisyan}},\ }\bibfield  {title} {\bibinfo {title} {Non-{Markovian} modeling of {Fermi}-{Bose} systems coupled to one or several {Fermi}-{Bose} thermal baths},\ }\href {https://doi.org/10.1103/PhysRevA.102.022209} {\bibfield  {journal} {\bibinfo  {journal} {Physical Review A}\ }\textbf {\bibinfo {volume} {102}},\ \bibinfo {pages} {022209} (\bibinfo {year} {2020})},\ \bibinfo {note} {publisher: American Physical Society}\BibitemShut {NoStop}%
\bibitem [{\citenamefont {Park}\ \emph {et~al.}(2024)\citenamefont {Park}, \citenamefont {Huang}, \citenamefont {Zhu}, \citenamefont {Yang}, \citenamefont {Chan},\ and\ \citenamefont {Lin}}]{park_quasi-lindblad_2024}%
  \BibitemOpen
  \bibfield  {author} {\bibinfo {author} {\bibfnamefont {G.}~\bibnamefont {Park}}, \bibinfo {author} {\bibfnamefont {Z.}~\bibnamefont {Huang}}, \bibinfo {author} {\bibfnamefont {Y.}~\bibnamefont {Zhu}}, \bibinfo {author} {\bibfnamefont {C.}~\bibnamefont {Yang}}, \bibinfo {author} {\bibfnamefont {G.~K.-L.}\ \bibnamefont {Chan}},\ and\ \bibinfo {author} {\bibfnamefont {L.}~\bibnamefont {Lin}},\ }\href {https://doi.org/10.48550/arXiv.2408.15529} {\bibinfo {title} {Quasi-{Lindblad} pseudomode theory for open quantum systems}} (\bibinfo {year} {2024}),\ \bibinfo {note} {arXiv:2408.15529 [cond-mat, physics:physics, physics:quant-ph]}\BibitemShut {NoStop}%
\bibitem [{\citenamefont {Arrigoni}\ \emph {et~al.}(2013)\citenamefont {Arrigoni}, \citenamefont {Knap}, \citenamefont {Linden},\ and\ \citenamefont {{Von Der Linden}}}]{arrigoni_nonequilibrium_2013}%
  \BibitemOpen
  \bibfield  {author} {\bibinfo {author} {\bibfnamefont {E.}~\bibnamefont {Arrigoni}}, \bibinfo {author} {\bibfnamefont {M.}~\bibnamefont {Knap}}, \bibinfo {author} {\bibfnamefont {W.~V.~D.}\ \bibnamefont {Linden}},\ and\ \bibinfo {author} {\bibfnamefont {W.}~\bibnamefont {{Von Der Linden}}},\ }\bibfield  {title} {\bibinfo {title} {Nonequilibrium {Dynamical} {Mean}-{Field} {Theory}: {An} {Auxiliary} {Quantum} {Master} {Equation} {Approach}},\ }\href {https://doi.org/10.1103/PhysRevLett.110.086403} {\bibfield  {journal} {\bibinfo  {journal} {Physical Review Letters}\ }\textbf {\bibinfo {volume} {110}},\ \bibinfo {pages} {086403} (\bibinfo {year} {2013})},\ \bibinfo {note} {publisher: American Physical Society},\ \Eprint {https://arxiv.org/abs/arXiv:1210.4167v2} {arXiv:arXiv:1210.4167v2} \BibitemShut {NoStop}%
\bibitem [{\citenamefont {Dorda}\ \emph {et~al.}(2014)\citenamefont {Dorda}, \citenamefont {Nuss}, \citenamefont {von~der Linden},\ and\ \citenamefont {Arrigoni}}]{dorda_auxiliary_2014}%
  \BibitemOpen
  \bibfield  {author} {\bibinfo {author} {\bibfnamefont {A.}~\bibnamefont {Dorda}}, \bibinfo {author} {\bibfnamefont {M.}~\bibnamefont {Nuss}}, \bibinfo {author} {\bibfnamefont {W.}~\bibnamefont {von~der Linden}},\ and\ \bibinfo {author} {\bibfnamefont {E.}~\bibnamefont {Arrigoni}},\ }\bibfield  {title} {\bibinfo {title} {Auxiliary master equation approach to nonequilibrium correlated impurities},\ }\href {https://doi.org/10.1103/PhysRevB.89.165105} {\bibfield  {journal} {\bibinfo  {journal} {Physical Review B}\ }\textbf {\bibinfo {volume} {89}},\ \bibinfo {pages} {165105} (\bibinfo {year} {2014})},\ \bibinfo {note} {publisher: American Physical Society}\BibitemShut {NoStop}%
\bibitem [{\citenamefont {Dorda}\ \emph {et~al.}(2015)\citenamefont {Dorda}, \citenamefont {Ganahl}, \citenamefont {Evertz}, \citenamefont {von~der Linden},\ and\ \citenamefont {Arrigoni}}]{dorda_auxiliary_2015}%
  \BibitemOpen
  \bibfield  {author} {\bibinfo {author} {\bibfnamefont {A.}~\bibnamefont {Dorda}}, \bibinfo {author} {\bibfnamefont {M.}~\bibnamefont {Ganahl}}, \bibinfo {author} {\bibfnamefont {H.~G.}\ \bibnamefont {Evertz}}, \bibinfo {author} {\bibfnamefont {W.}~\bibnamefont {von~der Linden}},\ and\ \bibinfo {author} {\bibfnamefont {E.}~\bibnamefont {Arrigoni}},\ }\bibfield  {title} {\bibinfo {title} {Auxiliary master equation approach within matrix product states: {Spectral} properties of the nonequilibrium {Anderson} impurity model},\ }\href {https://doi.org/10.1103/PhysRevB.92.125145} {\bibfield  {journal} {\bibinfo  {journal} {Physical Review B}\ }\textbf {\bibinfo {volume} {92}},\ \bibinfo {pages} {125145} (\bibinfo {year} {2015})},\ \bibinfo {note} {publisher: American Physical Society}\BibitemShut {NoStop}%
\bibitem [{\citenamefont {Schwarz}\ \emph {et~al.}(2016)\citenamefont {Schwarz}, \citenamefont {Goldstein}, \citenamefont {Dorda}, \citenamefont {Arrigoni}, \citenamefont {Weichselbaum},\ and\ \citenamefont {von Delft}}]{schwarz_lindblad-driven_2016}%
  \BibitemOpen
  \bibfield  {author} {\bibinfo {author} {\bibfnamefont {F.}~\bibnamefont {Schwarz}}, \bibinfo {author} {\bibfnamefont {M.}~\bibnamefont {Goldstein}}, \bibinfo {author} {\bibfnamefont {A.}~\bibnamefont {Dorda}}, \bibinfo {author} {\bibfnamefont {E.}~\bibnamefont {Arrigoni}}, \bibinfo {author} {\bibfnamefont {A.}~\bibnamefont {Weichselbaum}},\ and\ \bibinfo {author} {\bibfnamefont {J.}~\bibnamefont {von Delft}},\ }\bibfield  {title} {\bibinfo {title} {Lindblad-driven discretized leads for nonequilibrium steady-state transport in quantum impurity models: {Recovering} the continuum limit},\ }\href {https://doi.org/10.1103/PhysRevB.94.155142} {\bibfield  {journal} {\bibinfo  {journal} {Physical Review B}\ }\textbf {\bibinfo {volume} {94}},\ \bibinfo {pages} {155142} (\bibinfo {year} {2016})},\ \bibinfo {note} {publisher: American Physical Society}\BibitemShut {NoStop}%
\bibitem [{\citenamefont {Dorda}\ \emph {et~al.}(2017)\citenamefont {Dorda}, \citenamefont {Sorantin}, \citenamefont {von~der Linden},\ and\ \citenamefont {Arrigoni}}]{dorda_optimized_2017}%
  \BibitemOpen
  \bibfield  {author} {\bibinfo {author} {\bibfnamefont {A.}~\bibnamefont {Dorda}}, \bibinfo {author} {\bibfnamefont {M.}~\bibnamefont {Sorantin}}, \bibinfo {author} {\bibfnamefont {W.}~\bibnamefont {von~der Linden}},\ and\ \bibinfo {author} {\bibfnamefont {E.}~\bibnamefont {Arrigoni}},\ }\bibfield  {title} {\bibinfo {title} {Optimized auxiliary representation of non-{Markovian} impurity problems by a {Lindblad} equation},\ }\href {https://doi.org/10.1088/1367-2630/aa6ccc} {\bibfield  {journal} {\bibinfo  {journal} {New Journal of Physics}\ }\textbf {\bibinfo {volume} {19}},\ \bibinfo {pages} {063005} (\bibinfo {year} {2017})},\ \bibinfo {note} {publisher: IOP Publishing},\ \Eprint {https://arxiv.org/abs/1608.04632} {arXiv:1608.04632} \BibitemShut {NoStop}%
\bibitem [{\citenamefont {Fugger}\ \emph {et~al.}(2018)\citenamefont {Fugger}, \citenamefont {Dorda}, \citenamefont {Schwarz}, \citenamefont {Delft},\ and\ \citenamefont {Arrigoni}}]{fugger_nonequilibrium_2018}%
  \BibitemOpen
  \bibfield  {author} {\bibinfo {author} {\bibfnamefont {D.~M.}\ \bibnamefont {Fugger}}, \bibinfo {author} {\bibfnamefont {A.}~\bibnamefont {Dorda}}, \bibinfo {author} {\bibfnamefont {F.}~\bibnamefont {Schwarz}}, \bibinfo {author} {\bibfnamefont {J.~v.}\ \bibnamefont {Delft}},\ and\ \bibinfo {author} {\bibfnamefont {E.}~\bibnamefont {Arrigoni}},\ }\bibfield  {title} {\bibinfo {title} {Nonequilibrium {Kondo} effect in a magnetic field: auxiliary master equation approach},\ }\href {https://doi.org/10.1088/1367-2630/aa9fdc} {\bibfield  {journal} {\bibinfo  {journal} {New Journal of Physics}\ }\textbf {\bibinfo {volume} {20}},\ \bibinfo {pages} {013030} (\bibinfo {year} {2018})},\ \bibinfo {note} {publisher: IOP Publishing}\BibitemShut {NoStop}%
\bibitem [{\citenamefont {Schwarz}\ \emph {et~al.}(2018)\citenamefont {Schwarz}, \citenamefont {Weymann}, \citenamefont {von Delft},\ and\ \citenamefont {Weichselbaum}}]{schwarz_nonequilibrium_2018}%
  \BibitemOpen
  \bibfield  {author} {\bibinfo {author} {\bibfnamefont {F.}~\bibnamefont {Schwarz}}, \bibinfo {author} {\bibfnamefont {I.}~\bibnamefont {Weymann}}, \bibinfo {author} {\bibfnamefont {J.}~\bibnamefont {von Delft}},\ and\ \bibinfo {author} {\bibfnamefont {A.}~\bibnamefont {Weichselbaum}},\ }\bibfield  {title} {\bibinfo {title} {Nonequilibrium {Steady}-{State} {Transport} in {Quantum} {Impurity} {Models}: {A} {Thermofield} and {Quantum} {Quench} {Approach} {Using} {Matrix} {Product} {States}},\ }\href {https://doi.org/10.1103/PhysRevLett.121.137702} {\bibfield  {journal} {\bibinfo  {journal} {Physical Review Letters}\ }\textbf {\bibinfo {volume} {121}},\ \bibinfo {pages} {137702} (\bibinfo {year} {2018})},\ \bibinfo {note} {publisher: American Physical Society}\BibitemShut {NoStop}%
\bibitem [{\citenamefont {Chen}\ \emph {et~al.}(2019{\natexlab{a}})\citenamefont {Chen}, \citenamefont {Arrigoni},\ and\ \citenamefont {Galperin}}]{chen_markovian_2019}%
  \BibitemOpen
  \bibfield  {author} {\bibinfo {author} {\bibfnamefont {F.}~\bibnamefont {Chen}}, \bibinfo {author} {\bibfnamefont {E.}~\bibnamefont {Arrigoni}},\ and\ \bibinfo {author} {\bibfnamefont {M.}~\bibnamefont {Galperin}},\ }\bibfield  {title} {\bibinfo {title} {Markovian treatment of non-{Markovian} dynamics of open {Fermionic} systems},\ }\href {https://doi.org/10.1088/1367-2630/ab5ec5} {\bibfield  {journal} {\bibinfo  {journal} {New Journal of Physics}\ }\textbf {\bibinfo {volume} {21}},\ \bibinfo {pages} {123035} (\bibinfo {year} {2019}{\natexlab{a}})},\ \bibinfo {note} {publisher: IOP Publishing}\BibitemShut {NoStop}%
\bibitem [{\citenamefont {Chen}\ \emph {et~al.}(2019{\natexlab{b}})\citenamefont {Chen}, \citenamefont {Cohen},\ and\ \citenamefont {Galperin}}]{chen_auxiliary_2019}%
  \BibitemOpen
  \bibfield  {author} {\bibinfo {author} {\bibfnamefont {F.}~\bibnamefont {Chen}}, \bibinfo {author} {\bibfnamefont {G.}~\bibnamefont {Cohen}},\ and\ \bibinfo {author} {\bibfnamefont {M.}~\bibnamefont {Galperin}},\ }\bibfield  {title} {\bibinfo {title} {Auxiliary {Master} {Equation} for {Nonequilibrium} {Dual}-{Fermion} {Approach}},\ }\href {https://doi.org/10.1103/PhysRevLett.122.186803} {\bibfield  {journal} {\bibinfo  {journal} {Physical Review Letters}\ }\textbf {\bibinfo {volume} {122}},\ \bibinfo {pages} {186803} (\bibinfo {year} {2019}{\natexlab{b}})},\ \bibinfo {note} {publisher: American Physical Society}\BibitemShut {NoStop}%
\bibitem [{\citenamefont {Sorantin}\ \emph {et~al.}(2019)\citenamefont {Sorantin}, \citenamefont {Fugger}, \citenamefont {Dorda}, \citenamefont {von~der Linden},\ and\ \citenamefont {Arrigoni}}]{sorantin_auxiliary_2019}%
  \BibitemOpen
  \bibfield  {author} {\bibinfo {author} {\bibfnamefont {M.~E.}\ \bibnamefont {Sorantin}}, \bibinfo {author} {\bibfnamefont {D.~M.}\ \bibnamefont {Fugger}}, \bibinfo {author} {\bibfnamefont {A.}~\bibnamefont {Dorda}}, \bibinfo {author} {\bibfnamefont {W.}~\bibnamefont {von~der Linden}},\ and\ \bibinfo {author} {\bibfnamefont {E.}~\bibnamefont {Arrigoni}},\ }\bibfield  {title} {\bibinfo {title} {Auxiliary master equation approach within stochastic wave functions: {Application} to the interacting resonant level model},\ }\href {https://doi.org/10.1103/PhysRevE.99.043303} {\bibfield  {journal} {\bibinfo  {journal} {Physical Review E}\ }\textbf {\bibinfo {volume} {99}},\ \bibinfo {pages} {043303} (\bibinfo {year} {2019})},\ \bibinfo {note} {publisher: American Physical Society}\BibitemShut {NoStop}%
\bibitem [{\citenamefont {Fugger}\ \emph {et~al.}(2020)\citenamefont {Fugger}, \citenamefont {Bauernfeind}, \citenamefont {Sorantin},\ and\ \citenamefont {Arrigoni}}]{fugger_nonequilibrium_2020}%
  \BibitemOpen
  \bibfield  {author} {\bibinfo {author} {\bibfnamefont {D.~M.}\ \bibnamefont {Fugger}}, \bibinfo {author} {\bibfnamefont {D.}~\bibnamefont {Bauernfeind}}, \bibinfo {author} {\bibfnamefont {M.~E.}\ \bibnamefont {Sorantin}},\ and\ \bibinfo {author} {\bibfnamefont {E.}~\bibnamefont {Arrigoni}},\ }\bibfield  {title} {\bibinfo {title} {Nonequilibrium pseudogap {Anderson} impurity model: {A} master equation tensor network approach},\ }\href {https://doi.org/10.1103/PhysRevB.101.165132} {\bibfield  {journal} {\bibinfo  {journal} {Physical Review B}\ }\textbf {\bibinfo {volume} {101}},\ \bibinfo {pages} {165132} (\bibinfo {year} {2020})},\ \bibinfo {note} {publisher: American Physical Society}\BibitemShut {NoStop}%
\bibitem [{\citenamefont {Lotem}\ \emph {et~al.}(2020)\citenamefont {Lotem}, \citenamefont {Weichselbaum}, \citenamefont {von Delft},\ and\ \citenamefont {Goldstein}}]{lotem_renormalized_2020}%
  \BibitemOpen
  \bibfield  {author} {\bibinfo {author} {\bibfnamefont {M.}~\bibnamefont {Lotem}}, \bibinfo {author} {\bibfnamefont {A.}~\bibnamefont {Weichselbaum}}, \bibinfo {author} {\bibfnamefont {J.}~\bibnamefont {von Delft}},\ and\ \bibinfo {author} {\bibfnamefont {M.}~\bibnamefont {Goldstein}},\ }\bibfield  {title} {\bibinfo {title} {Renormalized {Lindblad} driving: {A} numerically exact nonequilibrium quantum impurity solver},\ }\href {https://doi.org/10.1103/PhysRevResearch.2.043052} {\bibfield  {journal} {\bibinfo  {journal} {Physical Review Research}\ }\textbf {\bibinfo {volume} {2}},\ \bibinfo {pages} {043052} (\bibinfo {year} {2020})},\ \bibinfo {note} {publisher: American Physical Society}\BibitemShut {NoStop}%
\bibitem [{\citenamefont {W\'ojtowicz}\ \emph {et~al.}(2020)\citenamefont {W\'ojtowicz}, \citenamefont {Elenewski}, \citenamefont {Rams},\ and\ \citenamefont {Zwolak}}]{wojtowicz_open_2020}%
  \BibitemOpen
  \bibfield  {author} {\bibinfo {author} {\bibfnamefont {G.}~\bibnamefont {W\'ojtowicz}}, \bibinfo {author} {\bibfnamefont {J.~E.}\ \bibnamefont {Elenewski}}, \bibinfo {author} {\bibfnamefont {M.~M.}\ \bibnamefont {Rams}},\ and\ \bibinfo {author} {\bibfnamefont {M.}~\bibnamefont {Zwolak}},\ }\bibfield  {title} {\bibinfo {title} {Open {System} {Tensor} {Networks} and {Kramers}’ {Crossover} for {Quantum} {Transport}},\ }\href {https://doi.org/10.1103/PhysRevA.101.050301} {\bibfield  {journal} {\bibinfo  {journal} {Physical review. A}\ }\textbf {\bibinfo {volume} {101}},\ \bibinfo {pages} {10.1103/PhysRevA.101.050301} (\bibinfo {year} {2020})}\BibitemShut {NoStop}%
\bibitem [{\citenamefont {Elenewski}\ \emph {et~al.}(2021)\citenamefont {Elenewski}, \citenamefont {W\'ojtowicz}, \citenamefont {Rams},\ and\ \citenamefont {Zwolak}}]{elenewski_performance_2021}%
  \BibitemOpen
  \bibfield  {author} {\bibinfo {author} {\bibfnamefont {J.~E.}\ \bibnamefont {Elenewski}}, \bibinfo {author} {\bibfnamefont {G.}~\bibnamefont {W\'ojtowicz}}, \bibinfo {author} {\bibfnamefont {M.~M.}\ \bibnamefont {Rams}},\ and\ \bibinfo {author} {\bibfnamefont {M.}~\bibnamefont {Zwolak}},\ }\bibfield  {title} {\bibinfo {title} {Performance of {Reservoir} {Discretizations} in {Quantum} {Transport} {Simulations}},\ }\href {https://doi.org/10.1063/5.0065799} {\bibfield  {journal} {\bibinfo  {journal} {The Journal of Chemical Physics}\ }\textbf {\bibinfo {volume} {155}},\ \bibinfo {pages} {124117} (\bibinfo {year} {2021})},\ \bibinfo {note} {arXiv:2105.01664 [cond-mat]}\BibitemShut {NoStop}%
\bibitem [{\citenamefont {Cirio}\ \emph {et~al.}(2023)\citenamefont {Cirio}, \citenamefont {Lambert}, \citenamefont {Liang}, \citenamefont {Kuo}, \citenamefont {Chen}, \citenamefont {Menczel}, \citenamefont {Funo},\ and\ \citenamefont {Nori}}]{cirio_pseudofermion_2023}%
  \BibitemOpen
  \bibfield  {author} {\bibinfo {author} {\bibfnamefont {M.}~\bibnamefont {Cirio}}, \bibinfo {author} {\bibfnamefont {N.}~\bibnamefont {Lambert}}, \bibinfo {author} {\bibfnamefont {P.}~\bibnamefont {Liang}}, \bibinfo {author} {\bibfnamefont {P.-C.}\ \bibnamefont {Kuo}}, \bibinfo {author} {\bibfnamefont {Y.-N.}\ \bibnamefont {Chen}}, \bibinfo {author} {\bibfnamefont {P.}~\bibnamefont {Menczel}}, \bibinfo {author} {\bibfnamefont {K.}~\bibnamefont {Funo}},\ and\ \bibinfo {author} {\bibfnamefont {F.}~\bibnamefont {Nori}},\ }\bibfield  {title} {\bibinfo {title} {Pseudofermion method for the exact description of fermionic environments: {From} single-molecule electronics to the {Kondo} resonance},\ }\href {https://doi.org/10.1103/PhysRevResearch.5.033011} {\bibfield  {journal} {\bibinfo  {journal} {Physical Review Research}\ }\textbf {\bibinfo {volume} {5}},\ \bibinfo {pages} {033011} (\bibinfo {year} {2023})},\ \bibinfo {note} {publisher: American Physical Society}\BibitemShut {NoStop}%
\bibitem [{\citenamefont {de~Vega}\ \emph {et~al.}(2015)\citenamefont {de~Vega}, \citenamefont {Schollw\"ock},\ and\ \citenamefont {Wolf}}]{de_vega_how_2015}%
  \BibitemOpen
  \bibfield  {author} {\bibinfo {author} {\bibfnamefont {I.}~\bibnamefont {de~Vega}}, \bibinfo {author} {\bibfnamefont {U.}~\bibnamefont {Schollw\"ock}},\ and\ \bibinfo {author} {\bibfnamefont {F.~A.}\ \bibnamefont {Wolf}},\ }\bibfield  {title} {\bibinfo {title} {How to discretize a quantum bath for real-time evolution},\ }\href {https://doi.org/10.1103/PhysRevB.92.155126} {\bibfield  {journal} {\bibinfo  {journal} {Physical Review B}\ }\textbf {\bibinfo {volume} {92}},\ \bibinfo {pages} {155126} (\bibinfo {year} {2015})},\ \bibinfo {note} {publisher: American Physical Society}\BibitemShut {NoStop}%
\bibitem [{\citenamefont {Hewson}(1993)}]{hewson_kondo_1993}%
  \BibitemOpen
  \bibfield  {author} {\bibinfo {author} {\bibfnamefont {A.~C.}\ \bibnamefont {Hewson}},\ }\href {https://doi.org/10.1017/CBO9780511470752} {\emph {\bibinfo {title} {The {Kondo} {Problem} to {Heavy} {Fermions}}}},\ Cambridge {Studies} in {Magnetism}\ (\bibinfo  {publisher} {Cambridge University Press},\ \bibinfo {address} {Cambridge},\ \bibinfo {year} {1993})\BibitemShut {NoStop}%
\bibitem [{\citenamefont {Besserve}\ \emph {et~al.}(2024)\citenamefont {Besserve}, \citenamefont {Ferrero},\ and\ \citenamefont {Ayral}}]{Besserve2024}%
  \BibitemOpen
  \bibfield  {author} {\bibinfo {author} {\bibfnamefont {P.}~\bibnamefont {Besserve}}, \bibinfo {author} {\bibfnamefont {M.}~\bibnamefont {Ferrero}},\ and\ \bibinfo {author} {\bibfnamefont {T.}~\bibnamefont {Ayral}},\ }\bibfield  {title} {\bibinfo {title} {Compact fermionic quantum state preparation with a natural-orbitalizing variational quantum eigensolving scheme}\ }\href {https://doi.org/10.48550/arXiv.2406.14170} {10.48550/arXiv.2406.14170} (\bibinfo {year} {2024}),\ \bibinfo {note} {arXiv:2406.14170 [quant-ph]}\BibitemShut {NoStop}%
\bibitem [{\citenamefont {Bravyi}\ and\ \citenamefont {Kitaev}(2002)}]{bravyi_fermionic_2002}%
  \BibitemOpen
  \bibfield  {author} {\bibinfo {author} {\bibfnamefont {S.~B.}\ \bibnamefont {Bravyi}}\ and\ \bibinfo {author} {\bibfnamefont {A.~Y.}\ \bibnamefont {Kitaev}},\ }\bibfield  {title} {\bibinfo {title} {Fermionic {Quantum} {Computation}},\ }\href {https://doi.org/10.1006/aphy.2002.6254} {\bibfield  {journal} {\bibinfo  {journal} {Annals of Physics}\ }\textbf {\bibinfo {volume} {298}},\ \bibinfo {pages} {210} (\bibinfo {year} {2002})}\BibitemShut {NoStop}%
\bibitem [{\citenamefont {Verstraete}\ and\ \citenamefont {Cirac}(2005)}]{verstraete_mapping_2005}%
  \BibitemOpen
  \bibfield  {author} {\bibinfo {author} {\bibfnamefont {F.}~\bibnamefont {Verstraete}}\ and\ \bibinfo {author} {\bibfnamefont {J.~I.}\ \bibnamefont {Cirac}},\ }\bibfield  {title} {\bibinfo {title} {Mapping local {Hamiltonians} of fermions to local {Hamiltonians} of spins},\ }\href {https://doi.org/10.1088/1742-5468/2005/09/P09012} {\bibfield  {journal} {\bibinfo  {journal} {Journal of Statistical Mechanics: Theory and Experiment}\ }\textbf {\bibinfo {volume} {2005}},\ \bibinfo {pages} {P09012} (\bibinfo {year} {2005})}\BibitemShut {NoStop}%
\bibitem [{\citenamefont {Endo}\ \emph {et~al.}(2020)\citenamefont {Endo}, \citenamefont {Kurata},\ and\ \citenamefont {Nakagawa}}]{endo2020}%
  \BibitemOpen
  \bibfield  {author} {\bibinfo {author} {\bibfnamefont {S.}~\bibnamefont {Endo}}, \bibinfo {author} {\bibfnamefont {I.}~\bibnamefont {Kurata}},\ and\ \bibinfo {author} {\bibfnamefont {Y.~O.}\ \bibnamefont {Nakagawa}},\ }\bibfield  {title} {\bibinfo {title} {Calculation of the {Green}'s function on near-term quantum computers},\ }\href {https://doi.org/10.1103/PhysRevResearch.2.033281} {\bibfield  {journal} {\bibinfo  {journal} {Physical Review Research}\ }\textbf {\bibinfo {volume} {2}},\ \bibinfo {pages} {033281} (\bibinfo {year} {2020})},\ \bibinfo {note} {publisher: American Physical Society}\BibitemShut {NoStop}%
\bibitem [{\citenamefont {Jamet}\ \emph {et~al.}(2023)\citenamefont {Jamet}, \citenamefont {Lenihan}, \citenamefont {Lindoy}, \citenamefont {Agarwal}, \citenamefont {Fontana}, \citenamefont {Martin},\ and\ \citenamefont {Rungger}}]{jamet2023}%
  \BibitemOpen
  \bibfield  {author} {\bibinfo {author} {\bibfnamefont {F.}~\bibnamefont {Jamet}}, \bibinfo {author} {\bibfnamefont {C.}~\bibnamefont {Lenihan}}, \bibinfo {author} {\bibfnamefont {L.~P.}\ \bibnamefont {Lindoy}}, \bibinfo {author} {\bibfnamefont {A.}~\bibnamefont {Agarwal}}, \bibinfo {author} {\bibfnamefont {E.}~\bibnamefont {Fontana}}, \bibinfo {author} {\bibfnamefont {B.~A.}\ \bibnamefont {Martin}},\ and\ \bibinfo {author} {\bibfnamefont {I.}~\bibnamefont {Rungger}},\ }\href {https://doi.org/10.48550/arXiv.2304.06587} {\bibinfo {title} {Anderson impurity solver integrating tensor network methods with quantum computing}} (\bibinfo {year} {2023}),\ \bibinfo {note} {arXiv:2304.06587 [quant-ph]},\ \Eprint {https://arxiv.org/abs/2304.06587} {arXiv:2304.06587} \BibitemShut {NoStop}%
\bibitem [{\citenamefont {Breuer}\ \emph {et~al.}(2007)\citenamefont {Breuer}, \citenamefont {Petruccione}, \citenamefont {Breuer},\ and\ \citenamefont {Petruccione}}]{breuerTheoryOpenQuantum2007}%
  \BibitemOpen
  \bibfield  {author} {\bibinfo {author} {\bibfnamefont {H.-P.}\ \bibnamefont {Breuer}}, \bibinfo {author} {\bibfnamefont {F.}~\bibnamefont {Petruccione}}, \bibinfo {author} {\bibfnamefont {H.-P.}\ \bibnamefont {Breuer}},\ and\ \bibinfo {author} {\bibfnamefont {F.}~\bibnamefont {Petruccione}},\ }\href@noop {} {\emph {\bibinfo {title} {The {Theory} of {Open} {Quantum} {Systems}}}}\ (\bibinfo  {publisher} {Oxford University Press},\ \bibinfo {address} {Oxford, New York},\ \bibinfo {year} {2007})\BibitemShut {NoStop}%
\bibitem [{\citenamefont {Rammer}(2007)}]{rammer_quantum_2007}%
  \BibitemOpen
  \bibfield  {author} {\bibinfo {author} {\bibfnamefont {J.}~\bibnamefont {Rammer}},\ }\href {https://doi.org/10.1017/CBO9780511618956} {\emph {\bibinfo {title} {Quantum {Field} {Theory} of {Non}-equilibrium {States}}}}\ (\bibinfo  {publisher} {Cambridge University Press},\ \bibinfo {address} {Cambridge},\ \bibinfo {year} {2007})\BibitemShut {NoStop}%
\bibitem [{\citenamefont {Boyd}\ and\ \citenamefont {Vandenberghe}(2004)}]{boyd_convex_2004}%
  \BibitemOpen
  \bibfield  {author} {\bibinfo {author} {\bibfnamefont {S.}~\bibnamefont {Boyd}}\ and\ \bibinfo {author} {\bibfnamefont {L.}~\bibnamefont {Vandenberghe}},\ }\href@noop {} {{\selectlanguage {English}\emph {\bibinfo {title} {Convex {Optimization}}}}},\ \bibinfo {edition} {1st}\ ed.\ (\bibinfo  {publisher} {Cambridge University Press},\ \bibinfo {address} {Cambridge, UK ; New York},\ \bibinfo {year} {2004})\BibitemShut {NoStop}%
\bibitem [{\citenamefont {Diamond}\ and\ \citenamefont {Boyd}(2016)}]{diamond2016cvxpy}%
  \BibitemOpen
  \bibfield  {author} {\bibinfo {author} {\bibfnamefont {S.}~\bibnamefont {Diamond}}\ and\ \bibinfo {author} {\bibfnamefont {S.}~\bibnamefont {Boyd}},\ }\bibfield  {title} {\bibinfo {title} {{CVXPY}: {A} {P}ython-embedded modeling language for convex optimization},\ }\href@noop {} {\bibfield  {journal} {\bibinfo  {journal} {Journal of Machine Learning Research}\ }\textbf {\bibinfo {volume} {17}},\ \bibinfo {pages} {1} (\bibinfo {year} {2016})}\BibitemShut {NoStop}%
\bibitem [{\citenamefont {Guo}\ and\ \citenamefont {Poletti}(2017)}]{guo_solutions_2017}%
  \BibitemOpen
  \bibfield  {author} {\bibinfo {author} {\bibfnamefont {C.}~\bibnamefont {Guo}}\ and\ \bibinfo {author} {\bibfnamefont {D.}~\bibnamefont {Poletti}},\ }\bibfield  {title} {\bibinfo {title} {Solutions for bosonic and fermionic dissipative quadratic open systems},\ }\href {https://doi.org/10.1103/PhysRevA.95.052107} {\bibfield  {journal} {\bibinfo  {journal} {Physical Review A}\ }\textbf {\bibinfo {volume} {95}},\ \bibinfo {pages} {052107} (\bibinfo {year} {2017})},\ \bibinfo {note} {publisher: American Physical Society}\BibitemShut {NoStop}%
\bibitem [{\citenamefont {Nielsen}\ and\ \citenamefont {Chuang}(2010)}]{Nielsen2010}%
  \BibitemOpen
  \bibfield  {author} {\bibinfo {author} {\bibfnamefont {M.~A.}\ \bibnamefont {Nielsen}}\ and\ \bibinfo {author} {\bibfnamefont {I.~L.}\ \bibnamefont {Chuang}},\ }\href {https://doi.org/10.1017/CBO9780511976667} {\emph {\bibinfo {title} {Quantum computation and quantum information}}}\ (\bibinfo  {publisher} {Cambridge University Press},\ \bibinfo {year} {2010})\BibitemShut {NoStop}%
\end{thebibliography}%
\end{document}